\documentclass[reprint,secnumarabic,amssymb, nobibnotes, aps, prd]{revtex4-1}

\usepackage[caption=false]{subfig}
\usepackage{graphicx}
\usepackage{amsmath}
\usepackage{amssymb}
\usepackage{braket}
\usepackage[table]{xcolor}
\usepackage{verbatim}
\usepackage{titlesec}
\usepackage[normalem]{ulem}
\newcommand{\fc}[1]{\textcolor{red}{#1}}

\begin{document}

	\title{Artificial Neural Networks as Trial Wave Functions for Quantum Monte Carlo}
	
	\author{Jan Kessler}%
	\author{Francesco Calcavecchia}
	\affiliation{Dynamics of Condensed Matter and Center for Sustainable Systems Design, Chair of Theoretical Chemistry, University of Paderborn, Warburger Str. 100, D-33098 Paderborn, Germany}
	\author{Thomas D. K\"uhne}
	\email{tkuehne@cp2k.org}
	\affiliation{Dynamics of Condensed Matter and Center for Sustainable Systems Design, Chair of Theoretical Chemistry, University of Paderborn, Warburger Str. 100, D-33098 Paderborn, Germany}
    \affiliation{Paderborn Center for Parallel Computing and Institute for Lightweight Design, University of Paderborn, Warburger Str. 100, D-33098 Paderborn, Germany}

	\begin{abstract}
	Inspired by the universal approximation theorem and widespread adoption of artificial neural network techniques in a diversity of fields, we propose feed-forward neural networks as a general purpose trial wave function for quantum Monte Carlo simulations of continous many-body systems. Whereas for simple model systems the whole many-body wave function can be represented by a neural network, the antisymmetry condition of non-trivial fermionic systems is incorporated by means of a Slater determinant. To demonstrate the accuracy of our trial wave functions, we have studied an exactly solvable model system of two trapped interacting particles, as well as the hydrogen dimer.
	\end{abstract}

	\maketitle

	\section{Introduction}
	
	Although the stationary Schr\"odinger equation describes the most basic form of quantum mechanics, finding accurate eigenstates and eigenvalues of the equation remains a formidable hard problem, at least when non-trivial realistic systems are concerned. Especially, including the important contribution of the electron correlation energy is computationally rather expensive when high accuracy is desired, which typically exhibits high-order polynomial or even exponential scaling effort with respect to the number of particles, thus rendering its application to complex systems infeasible.
	
	One class of methods for efficiently solving the many-body Schr\"odinger equation are variational Monte Carlo (VMC) methods \cite{MetropolisMC, VMC}, which are quantum Monte Carlo (QMC) schemes relying on optimizing a parametrized trial wave function according to the Rayleigh-Ritz variational principle and evaluating the inherently high-dimensional integrals by the means of importance-sampled Monte Carlo (MC) integration \cite{QMC, RevModPhys.73.33}. Typically, such an approach involves the postulation of problem-specific trial wave functions, in an attempt to reduce the number of required variational parameters based on physical intuition. The apparent downside of such an ansatz are the need of physical insights and the inflexibility of a specialized trial wave function, which is particularly problematic when, for example, phase transitions are meant to be investigated \cite{PhysRevE.90.053304, ZFNA}.
	
	Hence, in this work, we want to consider an overall similar, but conceptionally different approach. Instead of designing a specific trial wave function, we want to propose a novel class of very general trial wave functions built from feed-forward neural networks (FFNN) that is an artificial neural networks (ANN), wherein connections between the nodes do not form a cycle. The general idea is fueled by the so-called universal approximation theorem, which states that FFNNs can approximate any continuous function on a finite measure to arbitrary accuracy \cite{hornik_approximation_1991}. Furthermore, ANNs have already proven to be a successful ``black-box'' approach for a variety of problems in different fields and are therefore seen as particularly well-suited for dealing with high-dimensional inputs \cite{lecun_deep_2015}. In turn, however, the number of variational parameters within FFNNs grows large quickly and some need for intuition is shifted towards the question of how the particle coordinates are presented to the FFNN and what activation functions are to be used in the network.

	Even though, the first successful application of ANNs as wave functions for simple one-, two- and three-dimensional systems was already published more than two decades ago \cite{lagaris_artificial_1997}, the bulk of work in this direction, though only for lattice systems, is very recent \cite{carleo_solving_2017,gao_efficient_2017,saito_machine_2017,glasser_neural-network_2018,cai_approximating_2018,levine_bridging_2018}. Particular noteworthy approaches are based on ANN representations of restricted and deep Boltzmann machines \cite{carleo_constructing_2018}, as well as backflow transformations \cite{ruggeri_nonlinear_2018, PhysRevLett.122.226401}.
	Very recently, however, novel QMC methods for continuous space fermionic systems based on neural networks have appeared as preprints shortly after the present work \cite{KuhneArXiv,PfauPrint, NoePrint}. Whereas the latter two are genuine multi-reference approaches based on deep neural networks using multiple determinants, we aim at locating the fermionic ground state via FFNNs with just a single Slater determinant only. Whereas the former approaches have the potential to recover a slightly higher fraction of static correlation energy, the present method shares the simplicity and numerical efficiency of single-determinant Ans\"atze and has fewer parameters and is hence easier to optimize.
	To train the employed neural networks, we devise a gradient-based VMC optimization scheme. Both components of the present method promise a favorable computational scaling with respect to the dimensionality of the problem and allows for massively parallel execution, thus providing efficient use of modern supercomputers. 

    The remaining of the manuscript is organized as follows. First, in section~II, we briefly review the basic concepts of VMC and FFNNs, before introducing our novel neural network wave functions (NNWF). The application of the latter and the computational details are described in section~III. The corresponding results are presented in section~IV, followed by our conclusions in section~V.

	\section{Artifical Neural Network-Based \\ Quantum Monte Carlo}
	
	Before presenting our novel method for approximating ground state wave functions of continuous quantum systems by employing basic FFNNs as trial wave functions, we briefly review the main concepts of VMC that is based on the Rayleigh-Ritz variational principle, thus facilitating a systematic gradient-based strategy to train the eventual NNWFs.

	\subsection{Variational Monte Carlo} \label{ssec:vmc}
	
	From the non-relativistic time-independent spatial Schr\"odinger equation
	\begin{equation}
		\hat{H} \ket{\Psi(\mathbf{R})} = E \ket{\Psi(\mathbf{R})}
		\label{eq:schroedinger}
	\end{equation}
	we know that the exact spectrum $\Psi_{\alpha}(\mathbf{R})$ of stationary wave functions is given by the eigenstates of the Hamilton operator $\hat{H}$, which describes the physical system of concern. The corresponding eigenvalues $E_{\alpha}$ are the energies of the respective states. The state with the lowest energy $E_0$ is called ground state and labeled as $\Psi_0(\mathbf{R})$.
	
	However, for most realistic many-body Hamiltonians it is impossible to exactly solve Eq.~\ref{eq:schroedinger} and determine the exact spectrum of eigenstates, so the problem is to find accurate approximative representations of $\Psi_{\alpha}(\mathbf{R})$. To approximate the ground state $\Psi_0(\mathbf{R})$ in a systematic way, we make use of the Rayleigh-Ritz variational principle
	\begin{equation}
		\label{eq:rrvarp}
		E_0 \leq E_T = \frac{ \braket{\Psi_T|\hat{H}|\Psi_T} }{ \braket{\Psi_T|\Psi_T} } = \frac{ \int | \Psi_T |^2 \frac{\hat{H} \Psi_T}{\Psi_T} } {\int | \Psi_T |^2},
	\end{equation}
	where $\Psi_T = \Psi_T(\mathbf{R};\ \boldsymbol{\omega})$ is a parametrized trial wave function with parameters $\boldsymbol{\omega}$ and $E_T$ being the corresponding energy. Since $E_{T}$ is an upper bound of the true ground state energy $E_0$, the variational principle immediately suggests that we can optimize $\Psi_T$ by varying $\boldsymbol{\omega}$ so as to minimize $E_T$. To be specific, $\Psi_T(\mathbf{R})$ is a function of $D\times N$ particle coordinates, where $D$ is the dimensionality of the space and $N$ the number of particles. Computing arbitrary observables $O=\braket{\Psi_T|\hat{O}|\Psi_T}$ (e.g. $E_T$) that arise from such many-body states requires integrating over $\mathbb{R}^{D \times N}$ space, which results in solving rather high-dimensional integrals. Because of its favorable scaling with dimensionality, we employ MC integration for the numerical evaluation of these integrals.
	
	The rightmost term of Eq.~\ref{eq:rrvarp} immediately suggests how to compute the trial wave function's energy $E_T$ by importance sampled MC integration, in the fashion of a Metropolis-Hastings algorithm \cite{hastings_monte_1970}. In fact, given $\Psi_T(\mathbf{R})$, it is sufficient to compute the local energy
	\begin{equation}
		\label{eq:eloc}
		E_{loc}(\mathbf{R_i}) = \frac{\hat{H} \Psi_T(\mathbf{R_i})}{\Psi_T(\mathbf{R_i})}
	\end{equation}
	using coordinates $\mathbf{R_i}$ that are randomly sampled from the probability distribution $| \Psi_T( \mathbf{R} ) |^2$ and to average over them, i.e.
	\begin{equation}
		\label{eq:emc}
		E_T \approx E_{MC} = \frac{1}{N_{MC}} \sum_{i=1}^{N_{MC}} E_{loc}(\mathbf{R_i}).
	\end{equation}
  	In the equation above $\approx$ is used to indicate that the two values are equal within the statistical uncertainty due to the stochastic sampling. Notice that because Eq.~\ref{eq:emc} can be understood as a sum of independent summands, it can be evaluated in a massively parallel fashion by using multiple statistically independent random walks $\mathbf{R_i}$.
  	
  	This scheme is typically referred to as VMC method, where we optimize the parameters of the trial wave function by means of a gradient-based stochastic optimization algorithm, described in more detail in subsection~\ref{ssec:adam}.

	\subsection{Feed-Forward Neural Network Wave Functions}
	
	In the following, we describe how the specific NNWFs used in this work are constructed. For that purpose, we present multiple possible formulations of such wave functions, going from a straightforward and agnostic ansatz to more sophisticated and/or problem-specific variants.
	\label{ssec:ffnn}
	\begin{figure}
		\includegraphics[scale=0.4]{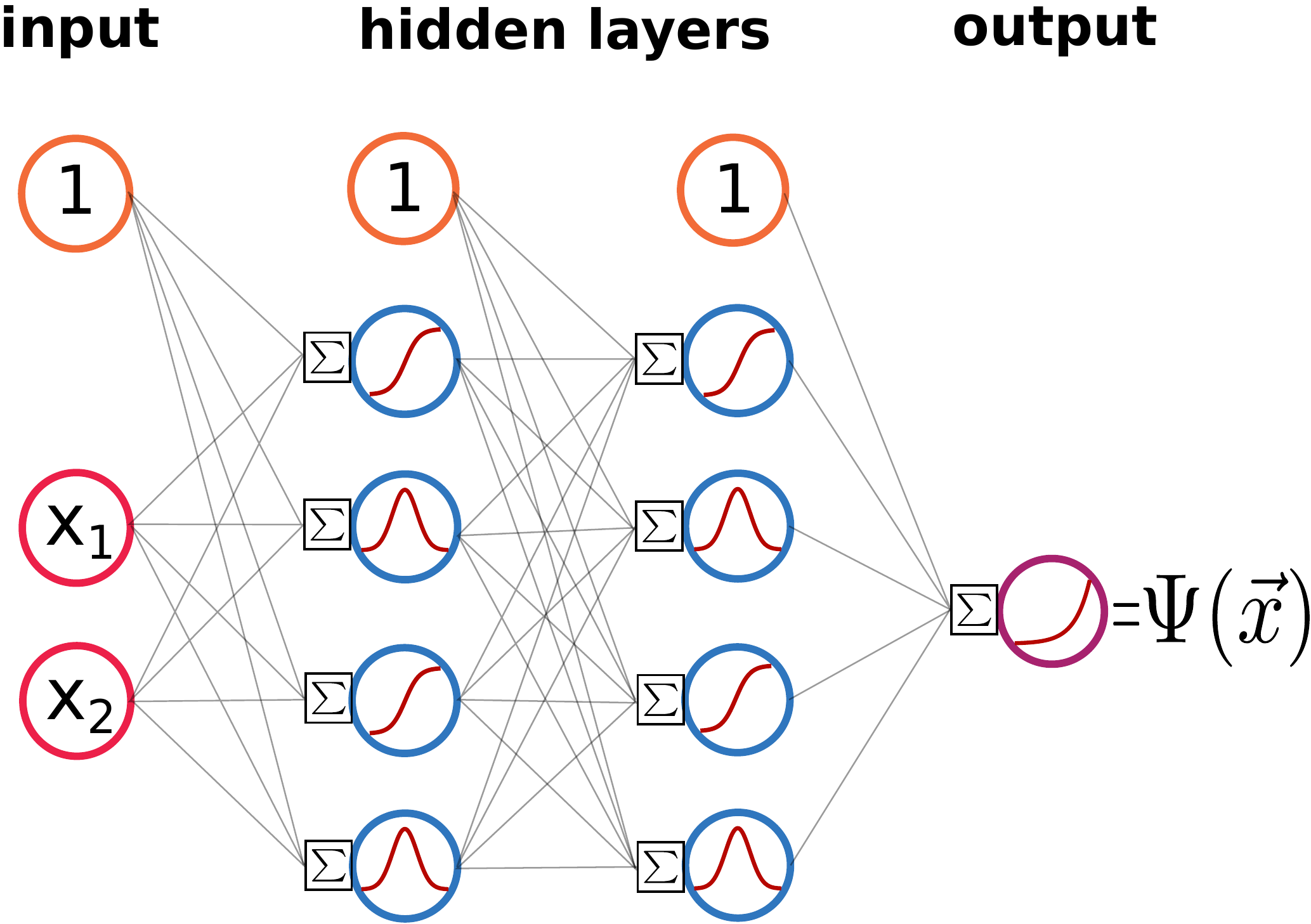}
		\caption{Illustration of a feed-forward NNWF with two input coordinates, two hidden layers, $4+1$ hidden units each and an exponential output activation function. The input units are depicted as red circles, offset units are orange, hidden units blue and output units purple, respectively. The graphs inside the circles represent the activation functions. The grey lines denote the unidirectional links between units, transferring information from left to right.} \label{fig:nnwf}
	\end{figure}
	More precisely, we describe how we construct trial wave functions for VMC simulations from basic FFNNs. The structure of such a network is illustrated in Fig.~\ref{fig:nnwf}. The employed ANNs can be divided into input, hidden and output layers, where each layer itself consists of multiple units, typically called artificial neurons or just units. Every such unit in the network signals a value to at least one other unit in the ``next'' layer via unidirectional connections, leading to a flow of information from the input (particle coordinates $\mathbf{R}$) to the output layer (wave function value $\Psi_T(\mathbf{R})$). 
	
	In the input layer, $D \times N$ input units are utilized to signal the coordinates $\mathbf{R}$ to all units of the next layer, the first hidden layer. Additionally, there is one offset unit, signaling a static $1$ in our case. Such an offset unit is present in every hidden layer as well. In the following hidden and output layers, the non-offset units are characterized by a propagation function $p(\mathbf{x};\ \boldsymbol{\omega})$ and an activation function $a(p)$. For these units the employed propagation function is a weighted sum of outputs $x_i$ of all units from the previous layer, i.e.
	\begin{equation}
		p(\mathbf{x};\ \boldsymbol{\omega}) = \boldsymbol{\omega} \cdot \mathbf{x} = \sum_i \omega_i x_i,
	\end{equation}
	where $\omega_i$ are the weights associated with each incoming connection. These weights are the variational parameters of our trial wave function. The result of the propagation function is then used as an input for the unit's activation function $a(p)$, to compute the final output value of the neuron, which is again transmitted to the next layer (if present).
  	
  	While in principle most non-linear functions are usable as activation functions, after some exploration (see appendix) we decided to use the following two hidden unit activation functions in all of our simulations:
  	\begin{align}
		\bullet\ &\text{Tangent-sigmoid:} &a_{T}(p) &= \frac{2}{1+\exp(-2 p)}-1 \label{eq:tans_actf} \\
		\bullet\ &\text{Gaussian:} &a_{G}(p) &= \exp(-p^2) \label{eq:gauss_actf}
	\end{align}
	Notably, these two activation functions exhibit opposite symmetry and different non-linearity. Also note that they are mathematically smooth, unlike many other frequently used neural network activation functions (e.g. rectified linear functions). However, our selection is certainly just one possible choice out of many. For the output unit we employed the exponential activation function
	\begin{equation}
		a_E(p) = \exp(p),
		\label{eq:exp_actf}
	\end{equation}
	which 
	entails a large range of output values, while confining the output to strictly positive numbers.
	For detailed informations about results using different other activation functions, we refer to the appendix.
	
	\subsubsection{Non-symmetric NNWF}
	
	For a first straightforward NNWF approach, we can simply feed the raw particle coordinates $\mathbf{R}$ directly into the described network via the input layer and utilize the output of the network as the wave function $\Psi(\mathbf{R})$. For simplicity, using just a single hidden layer with $n_h$ units (including an offset), we can write such a non-symmetric wave function as
	\begin{align}
		\label{eq:simple_nnwf}
		\Psi_{NSNN}& (\mathbf{R}) = \\* 
		& a_{o} \left( \omega_{o,0} + \sum_{i=1}^{n_h} \omega_{o,i} a_{h,i} \left( \omega_{h, i, 0} + \sum_{j=1}^{D \times N} \omega_{h,i,j} R_j \right) \right), \nonumber
	\end{align}
	where $a_o$ is the chosen output activation function (i.e. $a_L$ (Eq.~\ref{eq:log_actf}) or $a_E$ (Eq.~\ref{eq:exp_actf})) and $a_{h,i}$ the selected hidden layer activation functions (i.e. $a_T$ (Eq.~\ref{eq:tans_actf}) and/or $a_G$ (Eq.~\ref{eq:gauss_actf})). The corresponding connection weights within the output and hidden units, which will be determined by the training process, are denoted as $\boldsymbol{\omega}_o$ and $\boldsymbol{\omega}_{h,i}$, respectively. 
	
	A NNWF constructed in this way is (given smooth activation functions) notably a smooth function, which would be expected of a physical wave function for a finite potential energy. However, it exhibits no guaranteed symmetry under particle exchange. Yet, if the physical system of concern consists of distinguishable particles, the latter property is not directly an issue. But, the implications for the application to indistinguishable particles requiring bosonic or fermionic symmetry needs to be discussed. For example, a spatial wave function for (spinless) bosons is required to be strictly symmetric under arbitrary permutations of particle coordinates. However, it is known that for systems with local and exchange symmetric Hamiltonians without degeneracy and real-valued eigenstates, there is no other state with a lower energy than that of the symmetric ground state \cite{nagaoka_symmetry_1968}. In this case, by minimizing the energy, we would expect $\Psi_{NSNN}(\mathbf{R})$ to approximate the exchange symmetric ground state, thus becoming an approximately symmetric wave function.
	
	Although this ansatz is rather simplistic, we still consider it as an instructive test, as it is the maximally agnostic or ``naive'' approach and especially because it does not entail any additional computational cost for the explicit symmetrization. 

	\subsubsection{(Anti-)Symmetrized NNWF}
	
	Even though, it appears to be a potential option to use simple coordinate-fed FFNNs directly as wave functions - without guaranteed exchange symmetry - to approximate certain symmetric ground states, success is impossible if an approximation to an antisymmetric ground state is needed. Since the antisymmetric ground state is of higher energy than the corresponding symmetric ground state, minimizing the energy of an non-symmetric NNWF is expected to yield an approximately symmetric state. Moreover, in the special case of strictly positive valued output activation functions, the non-symmetric NNWF cannot possibly learn an antisymmetric state, by definition.
	
	The direct way of creating guaranteed exchange symmetric or antisymmetric wave functions from asymmetric base functions would be to apply the symmetrization or antisymmetrization operators, i.e. 
	\begin{subequations}
	\begin{eqnarray}
		\Psi_{SNN}(\mathbf{R}) &=& \mathcal{S}\ \Psi_{NSNN}(\mathbf{R}), \\ 
		\text{where}~\mathcal{S} &=& \frac{1}{N!} \sum_{P \in S_N} \hat{P}
		\label{eq:psi_S}
	\end{eqnarray}
	\end{subequations}
        and 
	\begin{subequations}
	\begin{eqnarray}
		\Psi_{ASNN}(\mathbf{R}) = \mathcal{A}\ \Psi_{NSNN}(\mathbf{R}), \\
		\text{where}~\mathcal{A} = \frac{1}{N!} \sum_{P \in S_N} (-1)^{\pi} \hat{P},
		\label{eq:psi_AS}
	\end{eqnarray}
	\end{subequations}
	respectively, 
	with $S_N$ being the symmetric group of permutations $\hat{P}$ and $\pi$ denoting the parity of the individual permutation.
	
	For the two-particle systems concerned in this work, we can directly write the resulting NNWF as
	\begin{equation}
		\Psi_{SNN/ASNN}(\mathbf{R}) = \frac{1}{2} \left( \Psi_{NSNN}(R_1, R_2) \pm \Psi_{NSNN}(R_2, R_1) \right).
	\end{equation}
	Constructing the NNWF in this way guarantees the desired exchange symmetry. However, it must be noted that the symmetrization operators implies a factorial computational scaling with the number of particles, which is prohibitive for more than a few particles. Nevertheless, for the present two-particle applications, we have included the results obtained with this ansatz, as a reference for comparison.
	
	\subsubsection{Product NNWF}
	
	So far, we have described a general method to obtain (in principle) arbitrarily exact approximations of real-valued spatial ground state wave functions from first-principles by taking only exchange symmetry into account. No further considerations about the physical system of concern were implemented into the construction of our trial wave functions. 
	While such a general and agnostic method has its own merits, we would not want to stay away from more specialized variants, at least when they prove to be simple to realize and more efficient. 
	
	One possible idea in this direction is to utilize a $\Psi_{NSNN}(\mathbf{R})$ merely as a modulating  NN-based Jastrow factor together with an imposed problem-specific determinantal wave function $\Phi(\mathbf{R})$, to form a product NNWF 
	\begin{equation}
		\Psi_{PNN}(\mathbf{R}) = \Psi_{NSNN}(\mathbf{R}) \ \Phi(\mathbf{R}).
	\end{equation}
	The determinantal part $\Phi(\mathbf{R})$ of $\Psi_{PNN}(\mathbf{R})$ could in principle be freely chosen, but we expect better results when it is already similar to the ground state that is to be approximated. Furthermore, $\Phi(\mathbf{R})$ directly prescribes the sign of the full product NNWF $\Psi_{PNN}(\mathbf{R})$, given that a strictly positive valued $\Psi_{NSNN}(\mathbf{R})$ is used. Then $\Psi_{PNN}(\mathbf{R})$ has the same sign as $\Phi(\mathbf{R})$ for all $\mathbf{R}$ and is equal to zero only at the same nodal surface $\{\mathbf{R} : \Phi(\mathbf{R}) = 0 \}$. This property effectively restricts the set of wave functions that such a product NNWF can approximate. 
	
	To approximate bosonic ground states, we consider a symmetric Hartree-product of identical single-particle orbitals $\phi(\mathbf{r})$ as the determinantal part of the NNWF, leading to
	\begin{equation}
		\Psi_{BPNN}(\mathbf{R}) = \Psi_{NSNN}(\mathbf{R})\ \phi(\mathbf{r_1})\phi(\mathbf{r_2})\ldots,
		\label{eq:bpnnwf}
	\end{equation}
	where $\mathbf{r_i}$ depict the coordinates of all individual particles. We will refer to $\Psi_{BPNN}(\mathbf{R})$ as bosonic product NNWF.
	
	For fermionic ground states, however, we use an antisymmetric Slater determinant $\Phi_{SD}$ of given single-particle orbitals $\phi_i(\mathbf{r})$ as the determinantal part of the eventual NNWF, i.e.
	\begin{equation}
		\Psi_{FPNN}(\mathbf{R}) = \Psi_{NSNN}(\mathbf{R})\ \Phi_{SD}(\mathbf{R}).
        \label{eq:fpnnwf}
	\end{equation}
	We denote $\Psi_{FPNN}(\mathbf{R})$ as fermionic product NNWF. As already alluded to above, given a strictly positive non-symmetric NNWF, the nodes of $\Psi_{FPNN}(\mathbf{R})$ are defined by $\Phi_{SD}(\mathbf{R})$. This is to say that as long as $\Phi_{SD}(\mathbf{R})$ does not change during the optimization, the nodes are fixed. If the nodal surface should be variable instead, the considered orbitals are required to contain variational parameters to be optimized together with the FFNN weights. One known possibility in this direction is the use of backflow orbitals \cite{PhysRevB.58.6800, PhysRevE.68.046707, PhysRevE.74.066701, Azadi_2013, PhysRevB.91.115106, azadi_2017, PhysRevB.97.205428}. However, for the specific applications of the present work, it suffice to rely on a fixed node ansatz.
	
	We find it important to mention that although at this time we have no mathematical proof at hand which would guarantee that the variational principle holds for $\Psi_{FPNN}(\mathbf{R})$ when using a non-symmetric NNWF, i.e. that there is no asymmetric state reachable with a truly lower energy than the exactly antisymmetric state. Nevertheless, we decided to test the product NNWF formulation in our applications, because it promises to solve the ``antisymmetry problem'' at negligible computational cost compared to the previously discussed full antisymmetrization, while still retaining the simplicity of raw coordinates $\mathbf{R}$.
	
	\subsubsection{Symmetric-Featured NNWF}
	
	Another opportunity for improvement concerns the way information is fed to the actual FFNN. Until now, we simply used all $D \times N$ raw particle coordinates $\mathbf{R}$ as input, but this choice might not be ideal in all cases. For example, considering the exchange symmetry, it is easy to see that the raw coordinates contain some redundant information, since for the wave function it does not matter if particle $1$ is at $\mathbf{r}_1$ and particle $2$ at $\mathbf{r}_2$, or the other way around. Instead, using a symmetric representation of all particle coordinates would ensure the exchange symmetry of the NNWF and, at the same time, potentially increase its accuracy. In general, an inspection of the specific Hamiltonian might lead to more efficient representations of all particle coordinates. For more complex systems, however, multi-particle features that are invariant with respect to translational and rotational symmetries may be essential \cite{BehlerJCP, PhysRevB.81.184107, PhysRevB.87.184115}. 
	
	Here, we propose and implement the possibility of such representations by replacing the $D \times N$ particle coordinates $\mathbf{R}_j$ in Eq.~\ref{eq:simple_nnwf} with $n_f$ functions $f_j(\mathbf{R})$ (``features''), resulting in
	\begin{align}
		\label{eq:featured_nnwf}
		\Psi_{SFNN}& (\mathbf{R}) = \\* 
		& a_{o} \left( \omega_{o,0} + \sum_{i=1}^{n_h} \omega_{o,i} a_{h,i} \left( \omega_{h, i, 0} + \sum_{j=1}^{n_f} \omega_{h,i,j} f_j(\mathbf{R}) \right) \right) \nonumber
	\end{align}
	for a network with only a single hidden layer. 
	
	In this work, we want to confine ourselves to exchange symmetric sets of features, which we will introduce in section \ref{sec:applications}, and therefore refer to such NNWFs as symmetric-featured NNWFs. Notably, employing a symmetric-featured NNWF as the Jastrow part of a bosonic or fermionic product NNWF yields a strictly symmetric or antisymmetric wave function. Hence, we will refer to a symmetric-featured product NNWF as  bosonic/fermionic product NNWFs that are defined as
	\begin{equation}
		\Psi_{BP-SFNN}(\mathbf{R}) = \Psi_{SFNN}(\mathbf{R})\ \phi(\mathbf{r_1})\phi(\mathbf{r_2})\ldots,
		\label{eq:bpsfnnwf}
	\end{equation}
	or
	\begin{equation}
		\Psi_{FP-SFNN}(\mathbf{R}) = \Psi_{SFNN}(\mathbf{R})\ \Phi_{SD}(\mathbf{R}),
        \label{eq:fpsfnnwf}
	\end{equation}
	respectively.
	
	At this point we want to note that the set of $f_j(\mathbf{R})$ must be chosen with some care. If relevant information is lost here, the symmetric-featured NNWF might not be able to approximate the true ground state anymore, no matter how large the employed FFNN is. Also important, the continuity and differentiability of our symmetric-featured NNWF will depend on the respective properties of the functions $f_j(\mathbf{R})$. 
	In the following, we present explicit problem-specific realizations of features $f_j(\mathbf{R})$ in section~\ref{sec:applications}.
	
	An overview of all the introduced NNWF types can be found in Table~\ref{tab:nnwfs}.
	\begin{widetext}
	\begin{center}
	\begin{table}
	\begin{tabular}{ |l|c|c| }
    \hline
    NNWF Name & Acronym & Imposed / Expected Symmetry\\
    \hline
    Non-symmetric & NSNN & None / Bosonic\\
    Symmetrized & SNN  & Bosonic / Bosonic\\
    Antisymmetrized & ASNN & Fermionic / Fermionic\\
    Bosonic product & BPNN & None / Bosonic\\
    Fermionic product & FPNN & None / Fermionic\\
    Symmetric-featured & SFNN  & Bosonic / Bosonic\\
    Symmetric-featured bosonic product & BP-SFNN  & Bosonic / Bosonic\\
    Symmetric-featured fermionic product & FP-SFNN  & Fermionic / Fermionic\\
    \hline
    \end{tabular}
    \caption{\label{tab:nnwfs}An overview of NNWF types and their properties, where imposed symmetry refers to an exchange symmetry that is guaranteed by construction and expected symmetry is a an (approximate) exchange symmetry that is expected to emerge from energy minimization.}
    \end{table}
    \end{center}
	\end{widetext}
	
	\section{Applications}
	\label{sec:applications}
	
	\subsection{2-particle harmonic trap with soft-core interactions}
	
	As a first application of our NNWFs, we selected a toy model system of two particles in a one-dimensional harmonic potential that are interacting via a potential of finite-range and constant value. Such a system can be described by the Hamiltonian
	\begin{equation}
		\hat{H} = - \frac{\left( \partial_{x_1}^2 + \partial_{x_2}^2 \right)}{2} + \frac{x_1^2 + x_2^2}{2} + \mathcal{V}(x_1 - x_2),
		\label{eq:ham_harmtrap}
	\end{equation}
	where we are using atomic units, harmonic oscillator frequency of $1\ a.u.$ and the soft-core potential
	\begin{equation}
		\mathcal{V}(x) =
		\begin{cases}
		V, & \text{if } |x| < a\\
		0, & \text{if } |x| \ge a
		\end{cases}.
	\end{equation}
	This model is a strongly correlated system, but still was shown to be exactly solvable for both bosonic and fermionic symmetry \cite{koscik_exactly_2018}. The outlined recipe to compute the exact solution allows us to obtain exact energy eigenvalues and eigenstates to validate our own simulation results. The soft-core interactions between the particles can be either attractive or repulsive, and both the potential range and strength can be chosen freely, thus providing a rich testing environment for our various NNWFs.
	
	While the Hamiltonian of Eq.~\ref{eq:ham_harmtrap} is all that is required to begin our WF optimization of the simple $\Psi_{NSNN}(\mathbf{x})$ or (anti-)symmetrized $\Psi_{S/AS}(\mathbf{x})$, for the product NNWFs $\Psi_{BP}(\mathbf{x})$ and $\Psi_{FP}(\mathbf{x})$ single-particle orbitals are required. To that extent, we consider the non-interacting system (i.e. $V=0$), namely a harmonic oscillator, which has well-known single-particle eigenstates given by the Hermite functions $\phi_n$. Relevant for us are the first two orbitals
	\begin{equation}
		\phi_0 (x) = \pi^{-\frac{1}{4}} e^{-\frac{1}{2}x^2} \ , \ \phi_1 (x) = \sqrt{2} \ \pi^{-\frac{1}{4}} x \ e^{-\frac{1}{2}x^2}.
	\end{equation}
	From the first or both of these orbitals we build the determinantal parts of our bosonic or fermionic product NNWFs, as in Eq.~\ref{eq:bpnnwf} or \ref{eq:fpnnwf}, respectively.
	
	In the case of symmetric-featured NNWFs we employ
	\begin{equation}
		f_1 = x_1^2 + x_2^2 \ , \ f_2 = (x_1 - x_2)^2
	\end{equation}
	as input of the FFNN, instead of the raw coordinates $\mathbf{x}$. For the a Hamiltonian, all relevant information are retained in this representation, while it guarantees bosonic exchange symmetry, spatial symmetry corresponding to the Hamiltonian and is mathematically smooth. We use this representation both for our symmetric-featured simple NNWFs and for the Jastrow part of symmetric-featured bosonic/fermionic product NNWFs.
	
	\subsection{H$_2$ Molecule}
	\label{ssec:applications_h2}
	
	As a more realistic application we study the hydrogen dimer H$_2$ within the Born-Oppenheimer approximation, i.e. we consider the electronic Hamiltonian (in atomic units)
	\begin{align}
		\hat{H} =& - \frac{1}{2} \left( \nabla_{\mathbf{r}_1}^2 + \nabla_{\mathbf{r}_2}^2 \right) - \sum_{i,j=1}^2 \frac{1}{ |\mathbf{r}_i - \mathbf{R}_j| } \nonumber \\*
		&+ \frac{1}{ |\mathbf{r}_1 - \mathbf{r}_2| } + \frac{1}{ |\mathbf{R}_1 - \mathbf{R}_2| },
		\label{eq:ham_h2}
	\end{align}
	where $\mathbf{r}$ denote the variable electronic coordinates and $\mathbf{R}$ the static protonic coordinates.
	
	Although this system consists of electrons, which are fermions, its spatial wave function must be exchange symmetric whenever the spin wave function is antisymmetric (singlet state). In fact, the state with the lowest energy is a singlet and therefore requires an exchange symmetric spatial wave function. Hence, we may either directly employ a simple NNWF, or use our bosonic product NNWF. For the latter, we use as single-particle orbital the bonding molecular orbital
	\begin{equation}
		\phi(\mathbf{r}) = e^{-|\mathbf{r}-\mathbf{R_1}|} + e^{-|\mathbf{r}-\mathbf{R_2}|},
	\end{equation}
	where $\mathbf{r}$ are the single electron coordinates and $\mathbf{R_i}$ the two protonic coordinates.
	
	Unlike for the previously discussed harmonic soft-core trap, the ground state of the hydrogen molecule cannot be obtained analytically. Instead, the total energies, as computed by full configuration-interaction method, are used as a reference for comparison  \cite{pachucki_born-oppenheimer_2010}. 

    \subsection{Adam-based VMC Optimization}
	\label{ssec:adam}
	
	Since a NNWF may contain hundreds if not thousands of variational parameters, we aim to adjust them in an effort to approximate the true ground state wave function as closely as possible, via an algorithm. A notable complication are the unavoidable statistical errors present in both our cost function and their respective gradients. To deal with this high-dimensional stochastic optimization problem, we employ the popular algorithm Adam \cite{kingma_adam:_2014}. It is a gradient-based optimization algorithm suitable for noisy gradients that internally builds up first and second order weight momenta and provides an adaptive learning rate for each of our weights $\omega_i$. We employ the algorithm as suggested by its authors, including the usage of an exponential moving average to obtain the optimized weights (see section 7.2 in Ref.~\onlinecite{kingma_adam:_2014}), leading to a reduced influence of noise on the final result.
	
	As cost function for the optimization we compute the energy $E_{T}$ according to Eq.~\ref{eq:emc} and to prevent overfitting we add a (small) $L_2$-regularization term for the variational parameters, i.e.
	\begin{equation}
		C = E_{T} + \frac{\lambda_r}{n_{\omega}} \sum_{i=1}^{n_{\omega}} \omega_i^2,
	\end{equation}
	where $n_\omega$ is the number of parameters and $\lambda_r$ a factor to control the magnitude of the term.
	
	The gradient of the energy $E_{T}$ can be derived analytically as
	\begin{align}
		\partial_{\omega_i} E_{T} &= \partial_{\omega_i} \frac{\braket{\psi|H|\psi}}{\braket{\psi|\psi}} \nonumber \\*
		&= 2 \left( \frac{\braket{\psi|H\ \partial_{\omega_i} |\psi}}{\braket{\psi|\psi}} - E_T  \frac{\braket{\psi|\partial_{\omega_i} |\psi}}{\braket{\psi|\psi}} \right).
	\end{align}
	This means that in our MC integration scheme we can compute the gradients $G_i$ of our cost function $C$ as
	\begin{align}
		G_{MC, i} &= \frac{2}{N_{MC}} \left[ \sum^{N_{MC}} \left( E_{loc} \frac{ \partial_{\omega_i} \Psi }{\Psi} \right) - E_{MC} \sum^{N_{MC}} \frac{ \partial_{\omega_i} \Psi }{\Psi} \right] \nonumber \\*
		& + 2\ \frac{\lambda_r}{n_{\omega}} \omega_i,
	\end{align}
	where the summands of sums $\sum^{N_{MC}}$ are evaluated along the importance sampled random walk, as discussed in section \ref{ssec:vmc}.
	
	Throughout the present work, this Adam-based optimizer is used to minimize the energy of all here proposed FFNN-based trial wave functions for a given Hamiltonian. 
	
	\section{Computational Details}
	\label{sec:compdet}
	
	In all of our simulations, the following basic strategy is employed to optimize the NNWF:
	\begin{enumerate}
		\item Least-squares fit of the NNWF to an initial-guess solution
		\item Use of our Adam-based VMC scheme to iteratively optimize the NNWF 
		\item Identify the result with the lowest variational energy 
	\end{enumerate}
	First, the NNWF is initialized to a well-behaved state using a relatively fast least-squares fitting procedure before the actual VMC optimization is performed. Even though this procedure is not strictly necessary, it increases efficiency and stability of the subsequent VMC algorithm, especially when the chosen initial state is already close to the true ground state solution. Except for product-type NNWFs, a product of Gaussian single-particle orbitals is employed to perform this step. Such an initial state appears to be a straightforward first guess for systems with localized ground states. For product-type NNWFs, however, we skipped the first step and started with weights initialized to small random numbers. Note that for technical reasons we did not use any symmetrizing operators during the fitting procedure, even when they were used in the following VMC optimization.
	
	Starting with the pre-fitted FFNN to represent the NNWF, we conducted an Adam-based optimization, which will be described in the following, with parameters $\alpha = 0.002$, $\beta_1 = 0.9$, $\beta_2 = 0.9$, $\epsilon = 10^{-8}$ (notation as in Ref.~\onlinecite{kingma_adam:_2014}) and regularization factor $\lambda_r = 0.005$. The variational parameters of the NNWF were initally updated according to the Adam algorithm until the last 100 optimization steps yielded constant energy, with respect to integration errors. With the resulting averaged variational parameters, a final MC evaluation of the respective energy was carried out. Afterwards, the Adam-based VMC optimization procedure was restarted, beginning with the final result from the previous optimization run. 
	
	Eventually, out of the individual resulting energies of all chained optimization runs, the one with the lowest upper energy confidence bound
	\begin{equation}
		E_{ub} = E_{MC} + 2~\sigma_{MC},
	\end{equation}
	with $\sigma_{MC}$ being the integration error estimation, is selected. This is done, since the last optimization run is not necessarily the one with the lowest variational energy, due to the stochastic nature of the optimization. These NNWFs were then employed within extensive MC sampling runs to evaluate the energies and all other observables reported in this paper.
	
	For each optimization step, the energies and gradients were computed based on $4*10^6$, the final energy evaluation of each chained optimization run based on $8*10^7$ and the eventual observables based on $10^9$ MC iterations, which were distributed among 16 CPU cores in all cases. For the MC sampling, we used all-particle moves proposed from a uniform distribution, where the maximum step-size was calibrated to yield an acceptance probability of $\approx 50 \%$ during $2500$ equilibration steps per thread. 
	For better stability when using NNWFs, we confined the walkers within a periodic interval of length $-10 < x < 10\ \text{Bohr}$, which is large enough to still allow to properly simulate  non-periodic systems of interest.
	
	All FFNNs employed here contained two hidden layers, each with $12$ hidden units and $1$ offset unit. A comparison of results for different network sizes can be found in Fig.~\ref{fig:comp_units} within the appendix.
	To realize our NNWF optimization in practice, we wrote our own C++ libraries for all involved tasks. All relevant libraries are available in the DCM-UPB/NNVMC repository on GitHub (https://github.com/DCM-UPB/NNVMC). 


	\section{Results}
	\subsection{2-particle harmonic trap with soft-core interaction}
		\begin{figure}
		\subfloat[]{\includegraphics[scale=0.525]{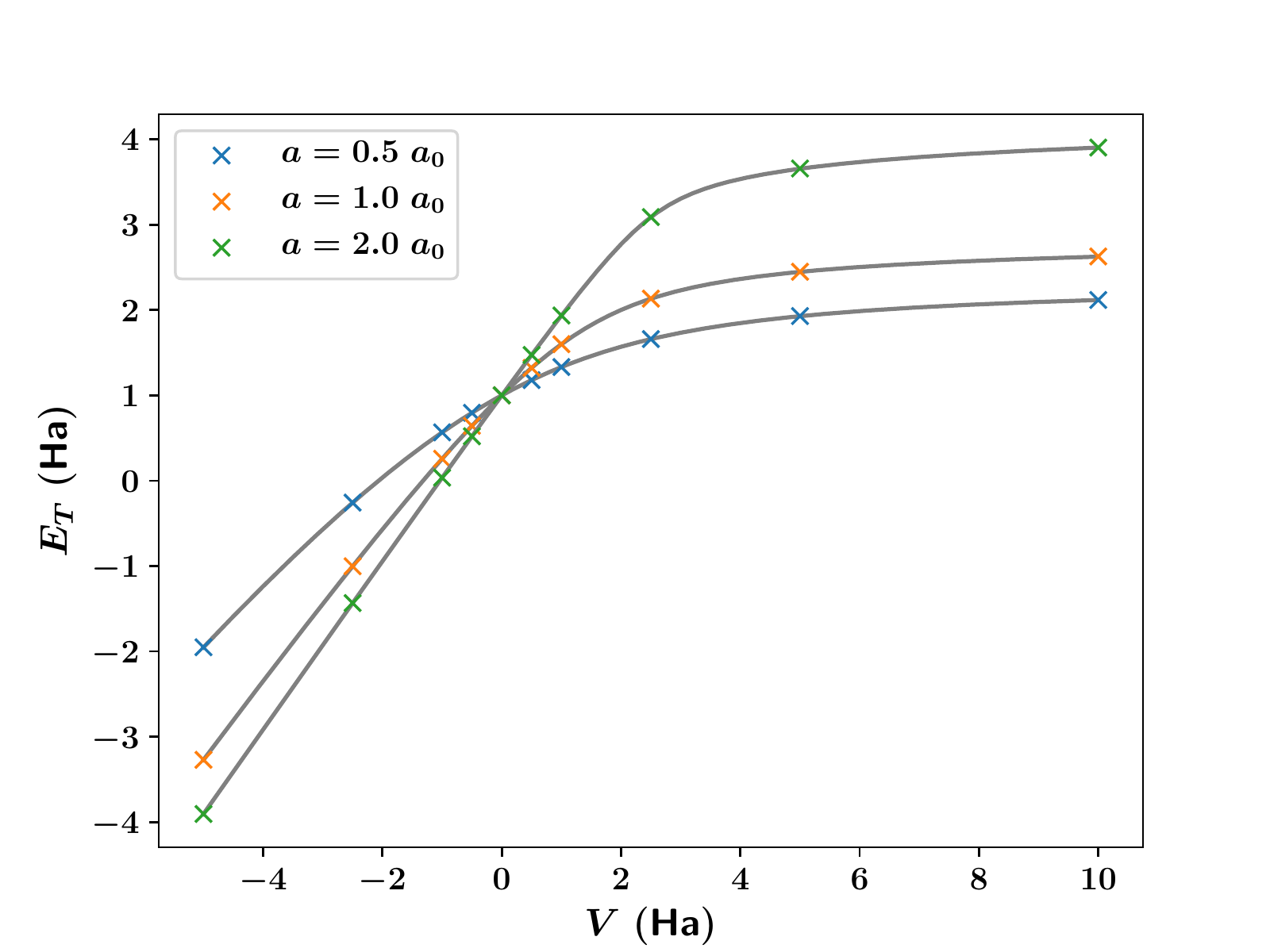} \label{subfig:harmtrap_energies_bose}} \hfill
		\subfloat[]{\includegraphics[scale=0.525]{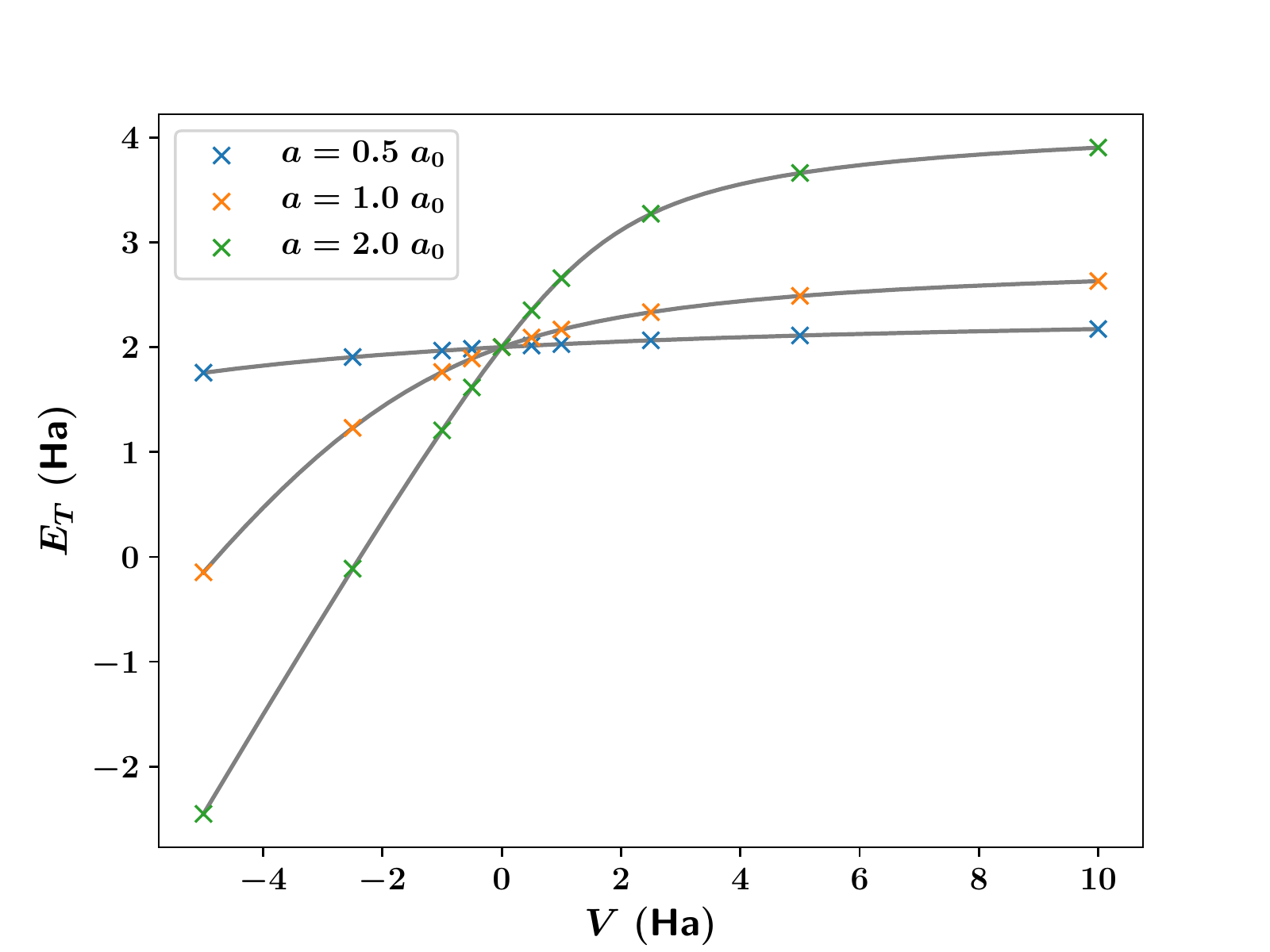} \label{subfig:harmtrap_energies_fermi}}
		\caption{Energies as obtained using (a) simple bosonic product and (b) simple fermionic product NNWFs, respectively. For comparison, the corresponding exact bosonic and fermionic ground state energies are shown (solid grey lines), for various potential ranges $a$ and potential strengths $V$ \cite{koscik_exactly_2018}.
		\label{fig:harmtrap_nosym_energies}}
	\end{figure}
	Using the previously described optimization scheme, we applied all of our NNWF methods to the soft-core harmonic trap system, for various choices of Hamiltonian parameters $a$ (potential range) and $V$ (potential strength). Exemplary, Fig.~\ref{fig:harmtrap_nosym_energies} shows the resulting energy values for simple bosonic (panel (a)) and fermionic (panel (b)) product NNWFs, in comparison with the exact bosonic and fermionic ground state energies computed based on the exact solutions presented in Ref.~\onlinecite{koscik_exactly_2018}. At least on the scale of these plots, the energies obtained from the NNWFs perfectly match the exact energies, for all considered Hamiltonian parameters. The same holds true for all other employed NNWFs. 
	
	For a more in-depth analysis of the results, we categorize all of our NNWF methods we have proposed here into two groups:
	\begin{enumerate}
		\item Non-symmetric NNWFs without strict exchange symmetry, i.e. plain NNWF, as well as simple bosonic and fermionic product NNWFs.
		\item (Anti-)Symmetric NNWFs with strict bosonic or fermionic exchange symmetry, i.e symmetrized and antisymmetrized NNWFs, symmetric-featured NNWF, as well as symmetric-featured bosonic and fermionc product NNWFs.
	\end{enumerate}
	For both groups we will always display the resulting energy residuals $E_T - E_0$, i.e. the difference between our VMC energy result $E_T$ and the exact ground state energy $E_0$. However, because the first group is technically prone to exhibit significant asymmetry, for these NNWFs we will also check positional observables and a measure of how well the desired exchange symmetry is realized.

	\subsubsection{Energy residuals}
	\begin{figure*}
		\subfloat[]{\includegraphics[scale=0.525]{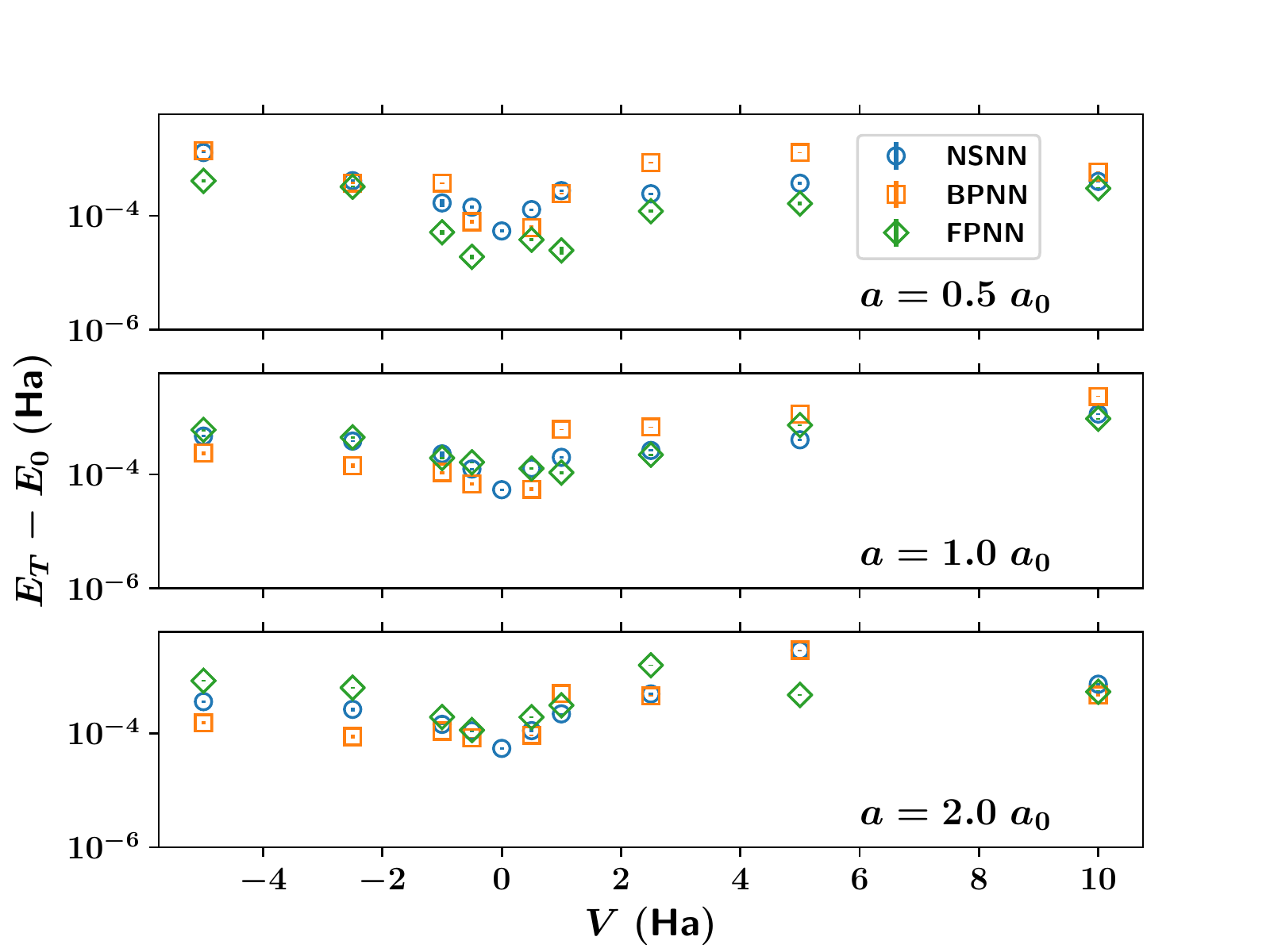} \label{subfig:harmtrap_nosym_residuals}} \hfill
		\subfloat[]{\includegraphics[scale=0.525]{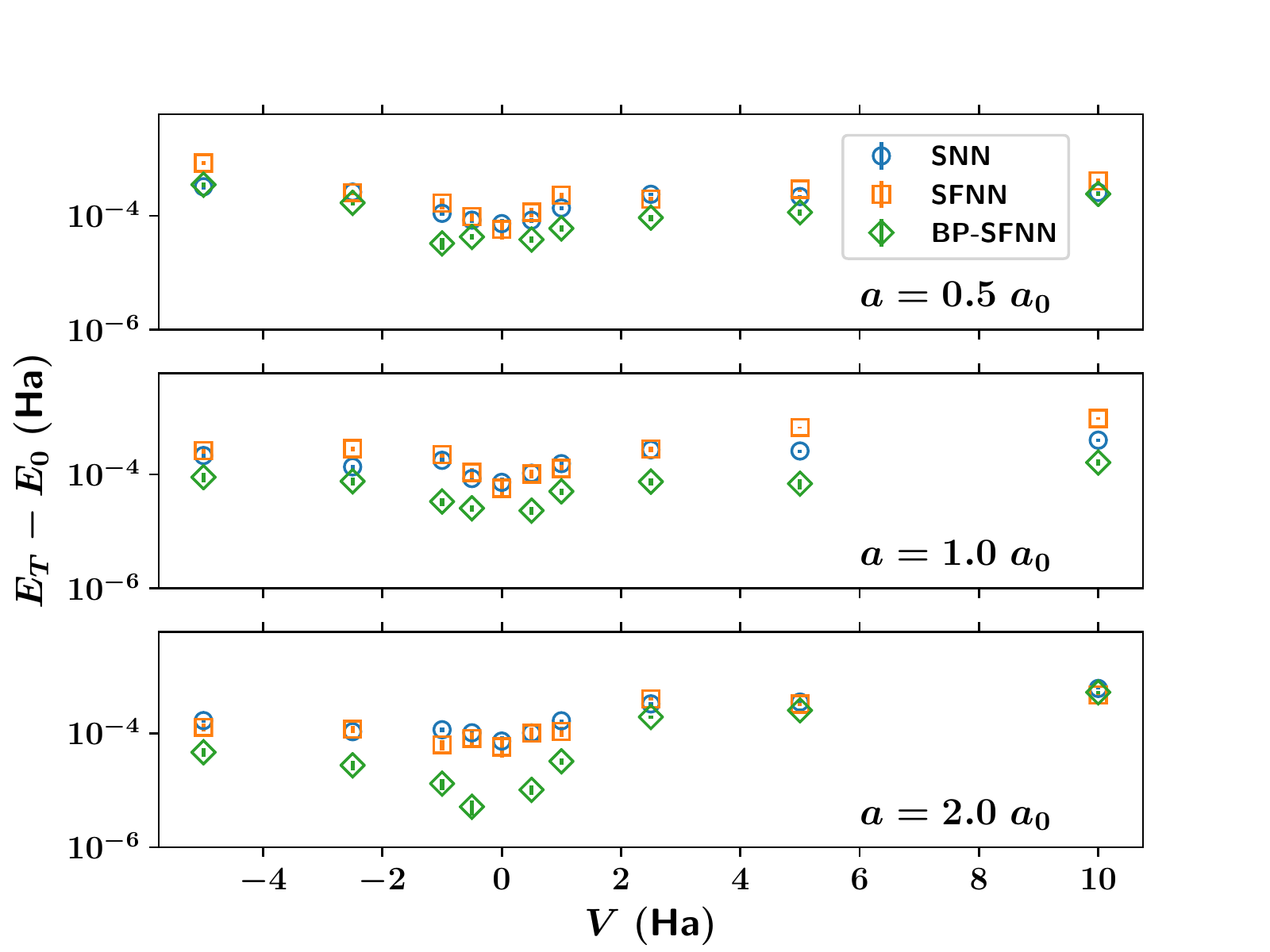} \label{subfig:harmtrap_sym_residuals}} \hfill
		\subfloat[]{\includegraphics[scale=0.525]{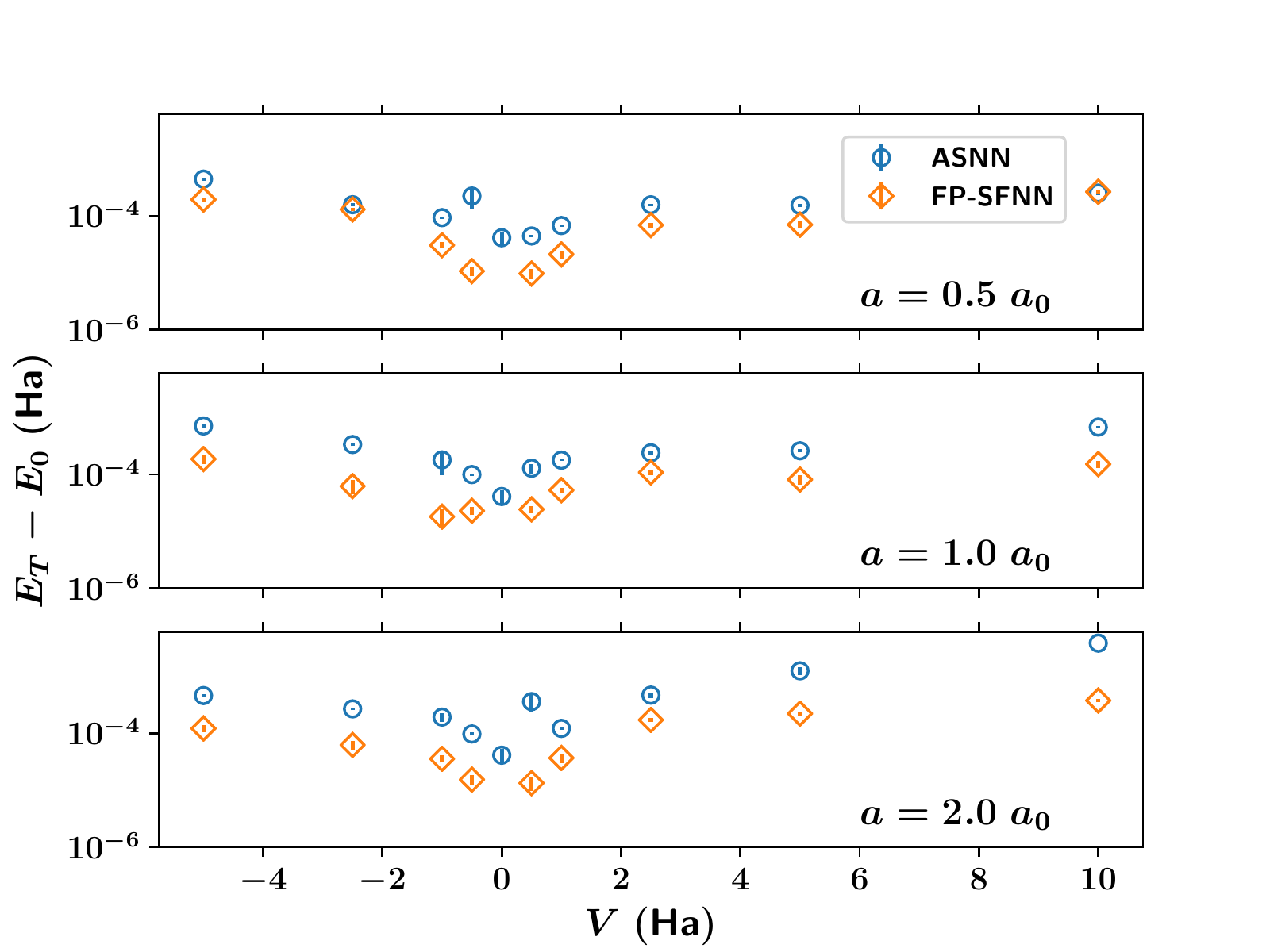} \label{subfig:harmtrap_antisym_residuals}} \hfill
		\subfloat[]{\includegraphics[scale=0.525]{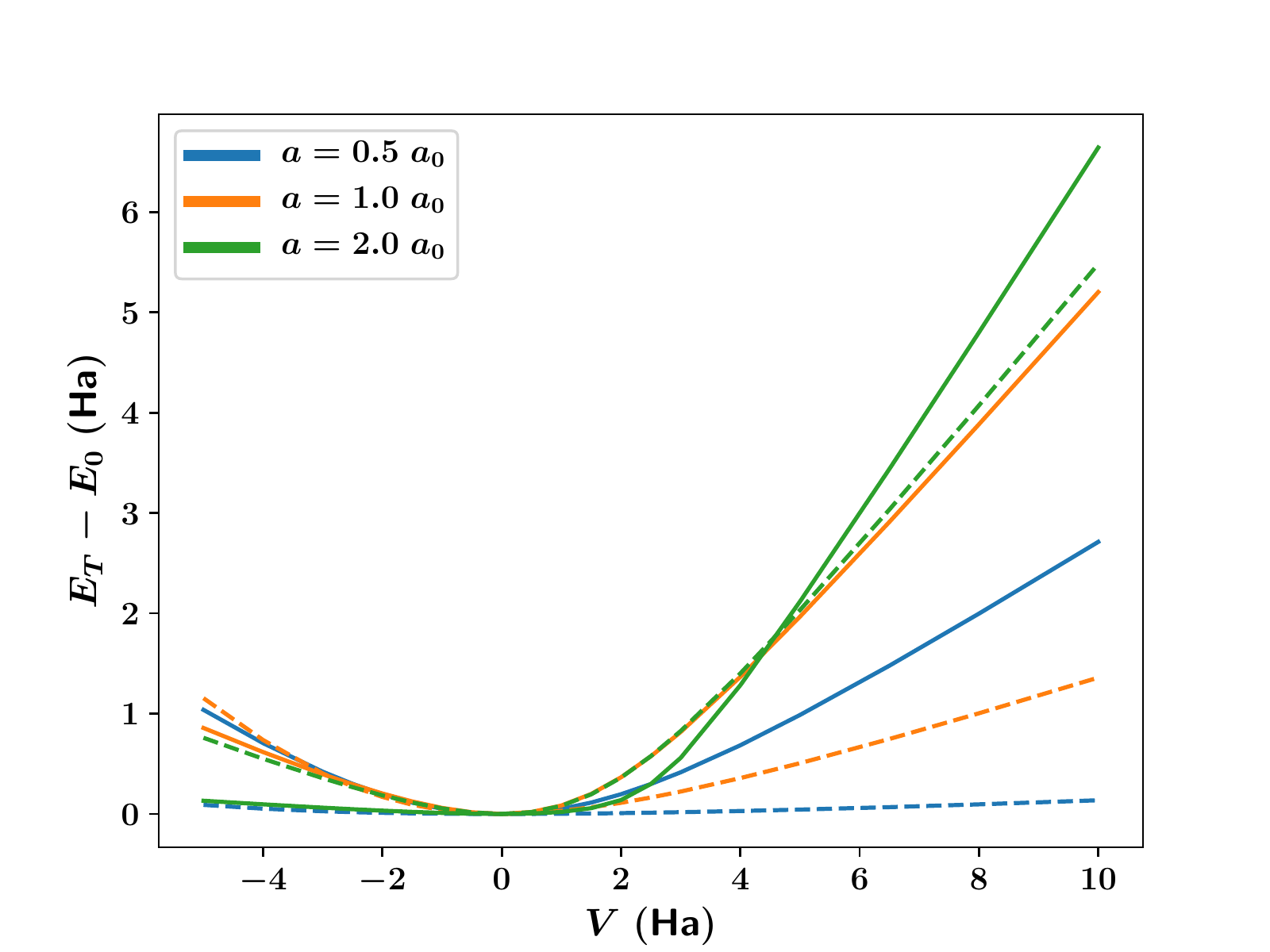} \label{subfig:harmtrap_hogs_residuals}}
		\caption{Energy residuals $E_T - E_0$, where $E_0$ is the exact solution  \cite{koscik_exactly_2018}, as obtained with (a) all not strictly exchange symmetric NNWFs, i.e. simple (NSNN), simple bosonic product (BPNN) and simple fermionic product (FPNN) NNWF, (b) all strictly exchange symmetric NNWFs, i.e. symmetrized (SNN), symmetric-featured (SFNN) and symmetric-featured bosonic product (BP-SFNN) NNWF, (c) all strictly exchange antisymmetric NNWFs, i.e. antisymmetrized (ASNN) and symmetric-featured fermionic product (FP-SFNN) NNWF, as well as (d) bosonic (solid lines) and fermionic (dashed lines) harmonic oscillator ground states (the determinantal parts of our product NNWFs), for various potential ranges $a$ and potential strengths $V$. Note the logarithmically scaled residual axis and that for product NNWFs, the trivial case of $V=0$ is left out.
		\label{fig:harmtrap_residuals}}
	\end{figure*}
    Beginning with the group of asymmetric NNWFs, we display their energy residuals in Fig.~\ref{subfig:harmtrap_nosym_residuals}. Note that the residual axis is scaled logarithmically to compensate for the large range of values. Furthermore, for any product-type NNWF, the interaction-free case $V=0$ is not shown, because the imposed determinantal part of our product NNWFs is already exactly the ground state of that system.
    
    In Fig.~\ref{subfig:harmtrap_nosym_residuals}, we are mainly looking for effects of using the simple bosonic-product NNWF over the simple non-symmetric NNWF. Overall, it appears that there is no clearly superior ansatz among the two. However, the product ansatz yields consistently similar or even lower energies, whenever the soft-core potential is attractive, or low by absolute value. In turn, it yields similar or higher energies for strong repulsive interactions. This behavior can be understood as a verification of our assumption that the product-type NNWF perform better when their imposed determinantal part is more similar to the true ground state. For our bosonic product NNWF we are using the bosonic ground state of a two-particle harmonic oscillator, which with some hand-waving can be said to be a first guess that tends to fit better when the inter-particle interaction is attractive or small and worse for strong repulsive interactions (also see Fig.~\ref{subfig:harmtrap_hogs_residuals}).
    
    Looking at the simple fermionic product NNWF, which uses the fermionic ground state of a two-particle harmonic oscillator, we get the opposite picture instead (as expected): It tends to perform better for repulsive interaction than for attractive interactions. Especially noteworthy is that the simple fermionic product NNWF predict accurate (residual $<10^{-3}$ Ha) fermionic energies without applying an antisymmetry operator, or using special coordinates to guarantee antisymmetry of the product.
    
    Now turning to the second group of wave functions, the energy residuals of all strictly exchange symmetric NNWFs is shown in Fig.~\ref{subfig:harmtrap_sym_residuals}, and those of all strictly exchange antisymmetric NNWFs in Fig.~\ref{subfig:harmtrap_antisym_residuals}. One main observation is that our symmetric-featured NNWF performs only marginally different to either simple or symmetrized NNWFs. However, if we use a network with symmetric feature input for the Jastrow part of the bosonic product ansatz, we obtain variationally energies that are consistently similar or even lower than those of any other bosonic NNWF ansatz that we investigated. Similarly, the symmetric-featured fermionic product NNWF performs throughout better than the all other fermionic NNWFs.
    
    Therefore, we found the product-type NNWFs to offer generally superior accuracy over the simple NNWF, at least when the imposed determinantal part of the wave function is already similar to the desired ground state. Furthermore, at least in combination with a product ansatz, we found the use of symmetrized features as neural network inputs to be beneficial, resulting in imporved variational energies. In any case, symmetrized feature inputs are always advantageous by guaranteeing the desired exchange symmetry or antisymmetry (when used in fermionic product NNWFs), without the factorial scaling computational cost as with symmetrized or antisymmetrized NNWFs.
    
    And finally, although there do exist differences in the accuracy of the various NNWF types, we find it noteworthy that although the same FFNN structure was employed in all these cases, the results are always satisfactory results without exception, even when just feeding raw coordinates to a FFNN and using the output as our wave function, which corresponds to our simple NNWF.
    
    \subsubsection{Asymmetry analysis}
		\begin{figure}
		\subfloat[]{\includegraphics[scale=0.525]{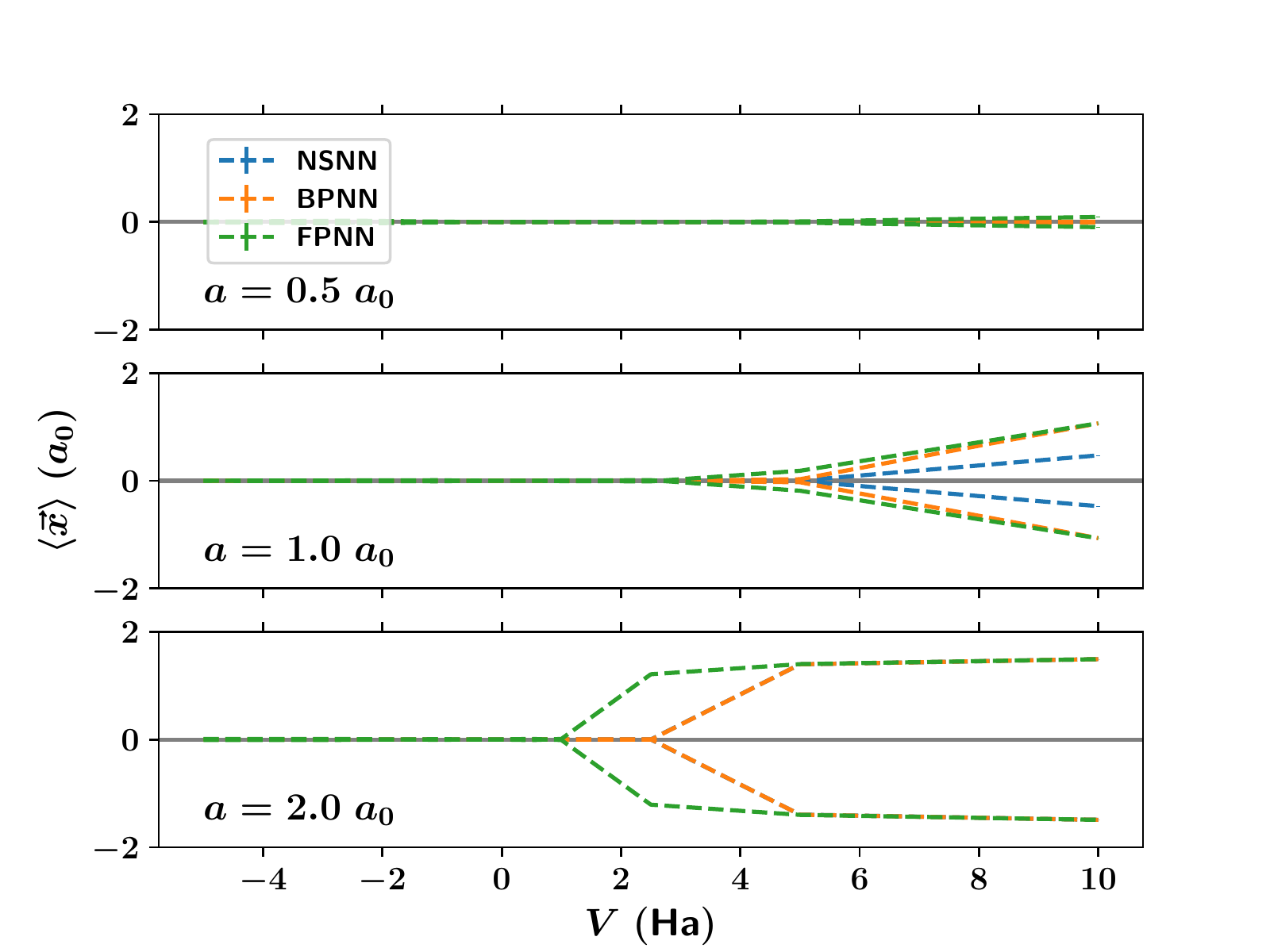} \label{subfig:harmtrap_position}} \hfill
		\subfloat[]{\includegraphics[scale=0.525]{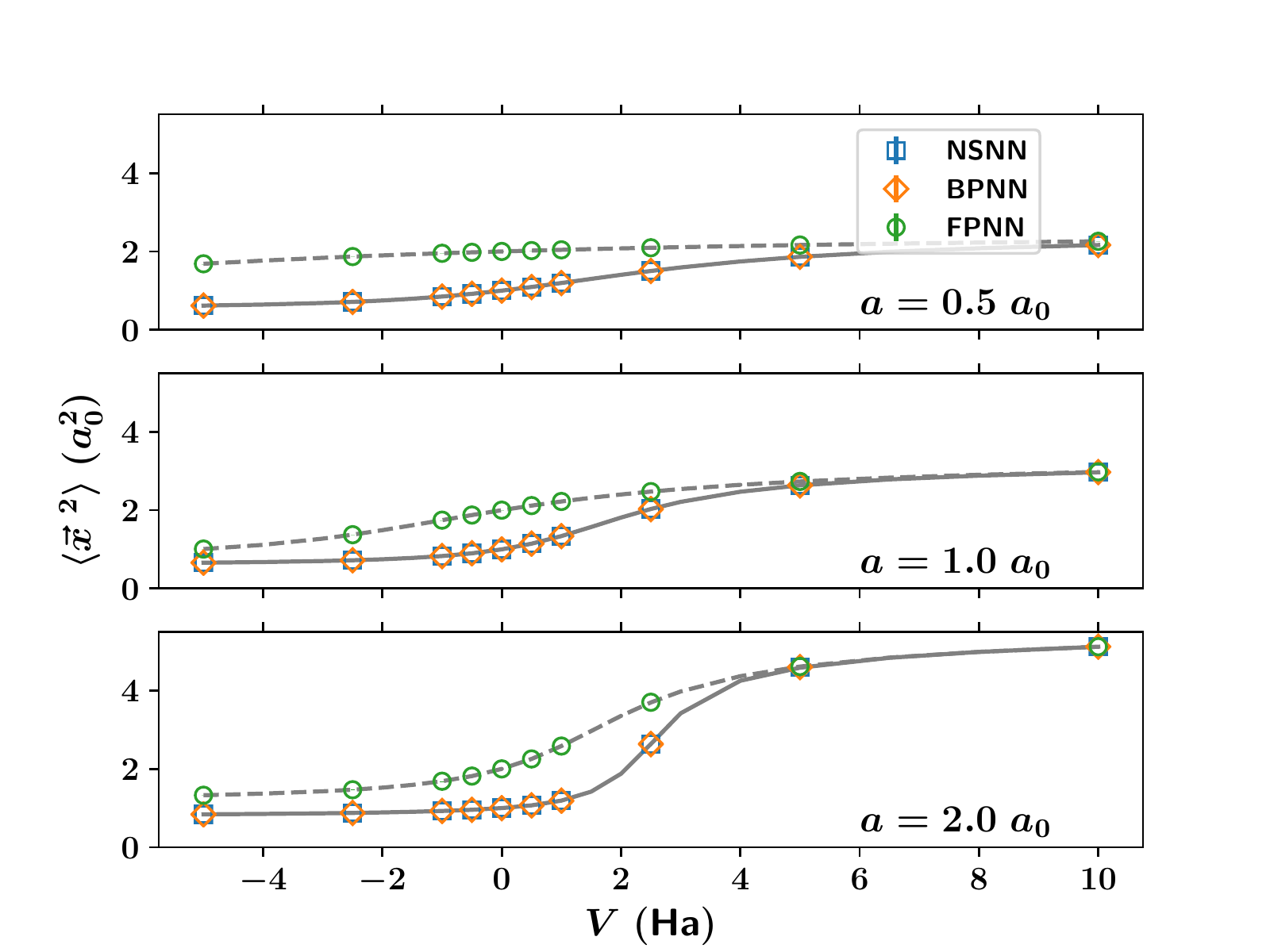} \label{subfig:harmtrap_positionsq}}
		\caption{Position expectation values (a) $\braket{x_1}$, $\braket{x_2}$ and (b) $\braket{\vec{x}^2} = \braket{x_1^2} + \braket{x_2^2}$, as obtained by our our non-symmetric NNWFs, i.e. simple NNWF (NSNN), simple bosonic product NNWF (BPNN) and simple fermionic product NNWF (FPNN), respectively. Grey lines depict the exact expectation values, with the solid line being for the bosonic and a dashed line for the fermionic case. 
		\label{fig:harmtrap_asymmetry}}
	\end{figure}
	\begin{figure}
		\includegraphics[scale=0.525]{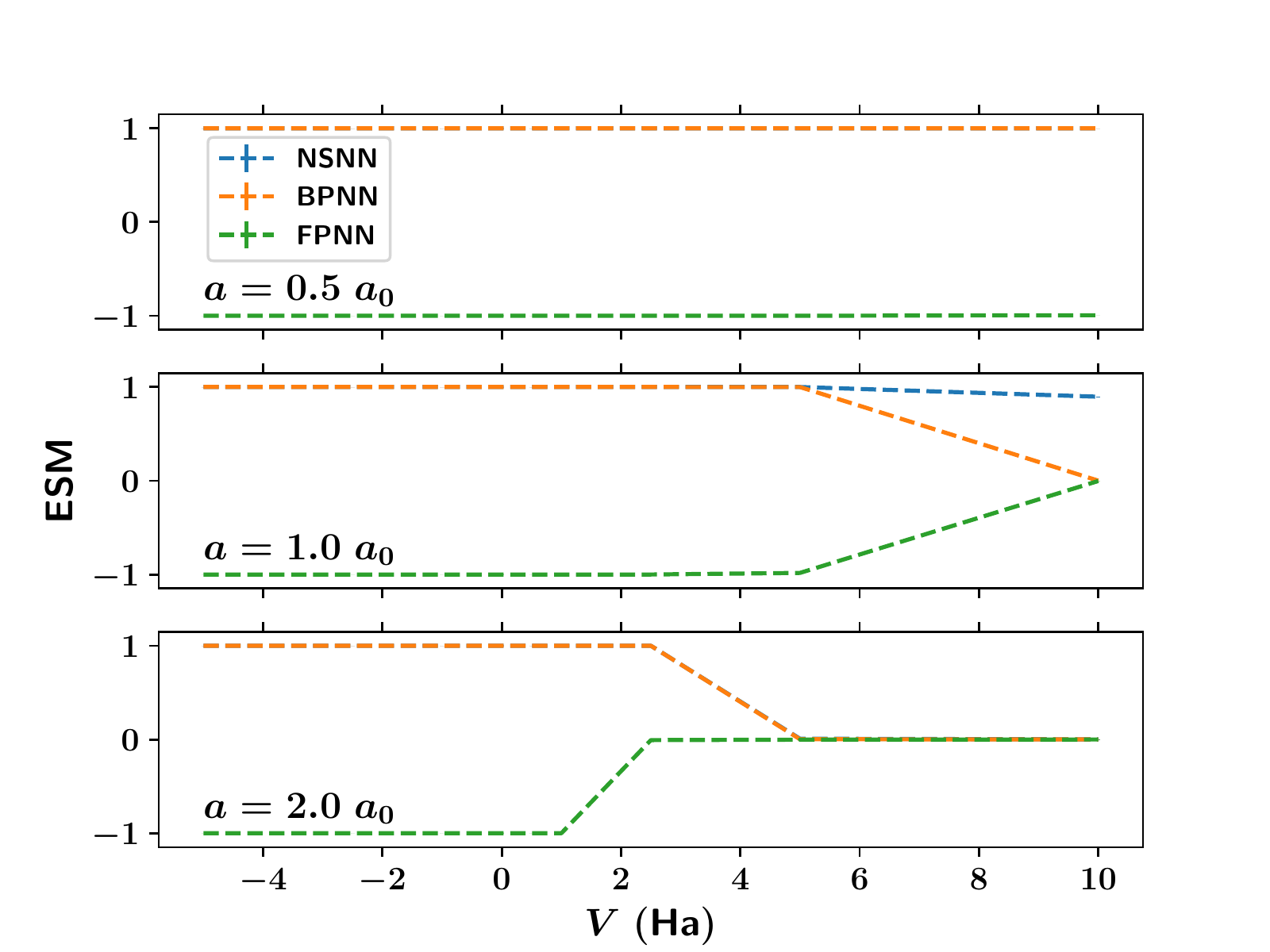}
		\caption{The ESM for all of our non-symmetric NNWFs, i.e. simple NNWF (NSNN), simple bosonic product NNWF (BPNN) and simple fermionic product NNWF (FPNN), applied to the soft-core harmonic trap system. An exactly symmetric/antisymmetric wave function would yield an ESM of +1/-1, respectively.
		\label{fig:harmtrap_esm}}
	\end{figure}
	\begin{figure}
		\includegraphics[scale=0.525]{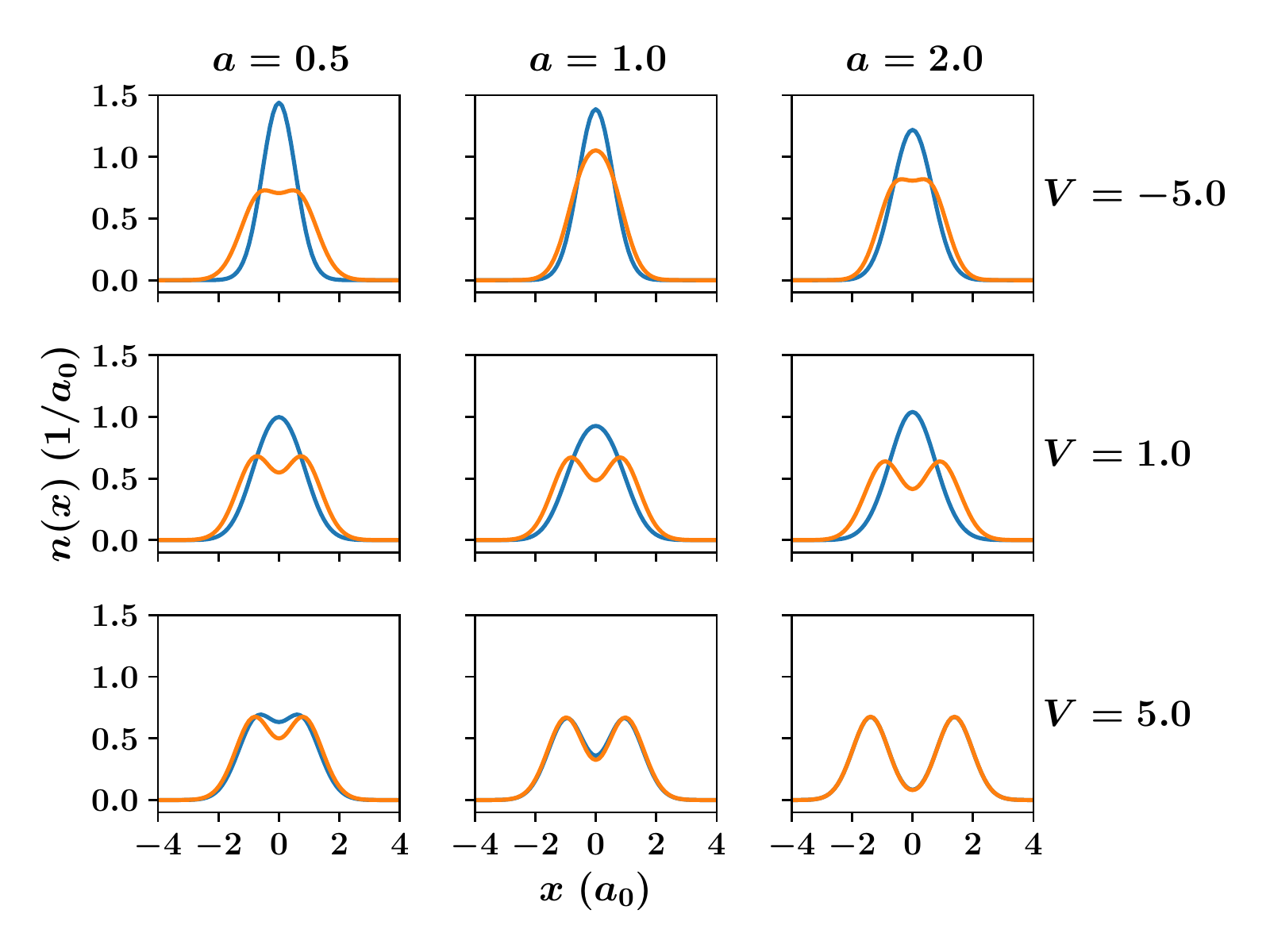}
		\caption{Density profiles $n(x)$ for the soft-core harmonic trap system, as obtained with simple bosonic (blue) and fermionic (orange) product NNWFs, respectively. All values are denoted in atomic units. The results are visually identical with the exact density profiles presented in Ref.~\onlinecite{koscik_exactly_2018}.
		\label{fig:harmtrap_density}}
	\end{figure}
	Although we have now verified that our NNWFs approximate the desired ground state energies, we have not shown yet whether or not the quality of approximation also holds for other observables. In particular, we want to validate the correct exchange symmetry and positional observables of the non-symmetric group of NNWFs. For this purpose, we computed the positional observables $\braket{\vec{x}}$ and $\braket{\vec{x}^2}$ (Fig.~\ref{fig:harmtrap_asymmetry}), as well as the exchange symmetry measure (Fig.~\ref{fig:harmtrap_esm}) 
	\begin{equation}
		ESM(\Psi) = \braket{\Psi(x_1, x_2)|\Psi(x_2,x_1)} / \braket{\Psi(\mathbf{x})|\Psi(\mathbf{x})},
		\label{eq:esm}
	\end{equation}
	which yields $1$ for a symmetric $\Psi$, $-1$ for an antisymmetric $\Psi$ and $0$ for a completely asymmetric $\Psi$. To complete the picture, we also computed the particle density $n(x) = \braket{\sum_i \delta(x_i -x)}$, as shown in Fig.~\ref{fig:harmtrap_density}.
	
	Looking at the results in Figs.~\ref{fig:harmtrap_asymmetry} and \ref{fig:harmtrap_esm}, respectively, it becomes immediately obvious that for certain Hamiltonian parameters $a \gg 0$ and $V \gg 0$, the observed exchange symmetry and position expectation values do not match the exact expectation values. Because we are considering indistinguishable particles and an external potential that is symmetric around the origin, we would expect both position expectation values $\braket{\vec{x}}$ to be identical and zero, regardless of Hamiltonian parameters. In contrast though, for the just mentioned choices for $a$ and $V$, we find the non-symmetric NNWFs to predict the two particles being located on opposite sides of the trap, i.e. $\braket{x_1} = - \braket{x_2} \neq 0$, and the symmetry measure is vanishing simultaneously. At the same time, the energy $E_T$ and other expectation values like $\braket{\vec{x}^2}$ and $n(x)$ remain close to the exact expectation. These properties indicate that our NNWF has actually learned a quasi-degenerate state of crystallized distinguishable particles, instead of the targeted bosonic or fermionic ground states. Notably, we observe this phenomenon only in a regime where bosonic and fermionc ground states itself become increasingly quasi-degenerate, which is also nicely illustrated by the density profiles in Fig.~\ref{fig:harmtrap_density}. The behavior of our non-symmetric NNWFs for large $V$ is fortunately not surprising, considering the analysis of the exact Hamiltonian eigenstates in the hard-core limit $V \rightarrow \infty$ \cite{koscik_exactly_2018}, where the states can be exactly constructed from single-particle orbitals located on the left and right sides of the trap.
	
	This is to say, we learned that if we chose to employ NNWFs without strict exchange symmetry to describe indistinguishable particles, the VMC optimization of our NNWF might prefer an asymmetric state of distinguishable particles whenever the corresponding energy is sufficiently close to the ground state energy. Hence, when for some reason using symmetric features is not an option and a non-symmetric NNWF must be used, one may check a diagnostic like Eq.~\ref{eq:esm} to verify the desired symmetry, if this is of practical relevance. However, if exchange symmetry of the NNWF can be guaranteed, e.g. by providing the used FFNN with an exchange symmetric input, the phenomenon discussed in this subsection is directly avoided.
	
	\subsection{Hydrogen dimer}
	\begin{figure}
		\subfloat[]{\includegraphics[scale=0.525]{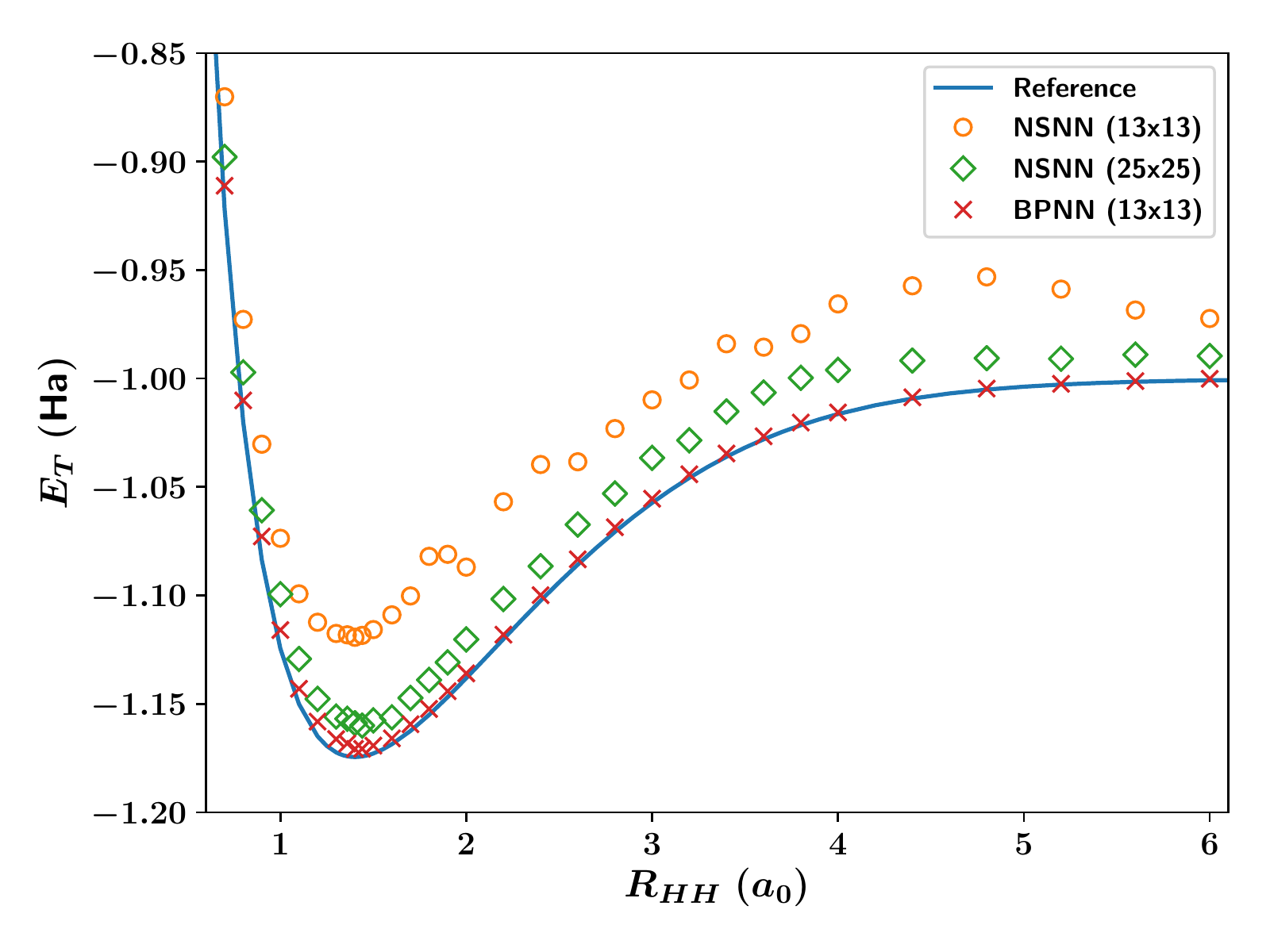} \label{subfig:h2_energies}} \hfill
		\subfloat[]{\includegraphics[scale=0.525]{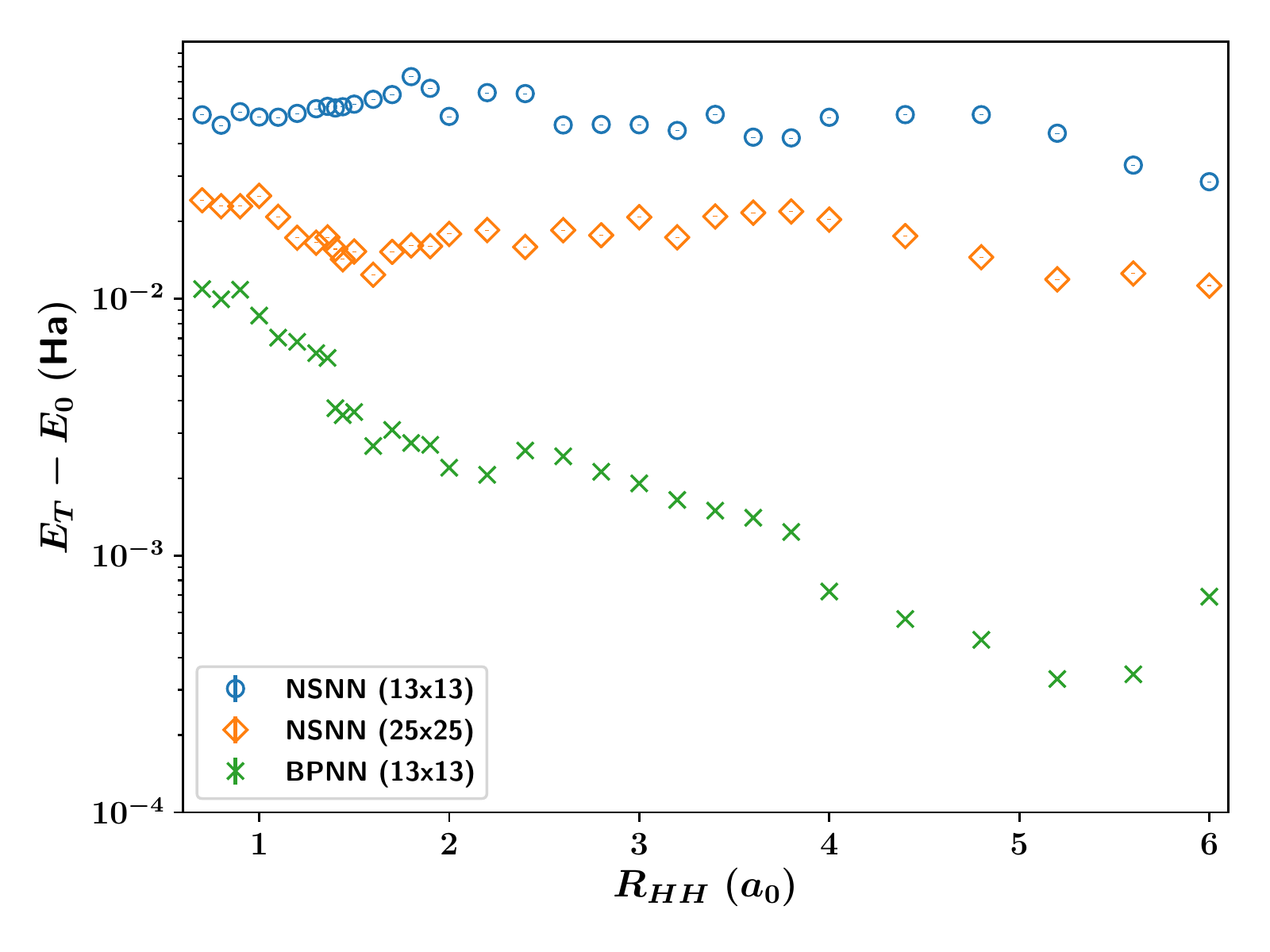} \label{subfig:h2_residuals}}
		\caption{(a) Energies and (b) energy residuals of the H$_2$ molecule as a function of separation distance $R_{HH}$ within the Born-Oppenheimer approximation, as obtained by our simple NNWF (NSNN) and simple bosonic product NNWF (BPNN), with $13$ and $25$ hidden units (including $1$ offset unit) per hidden layer.
		\label{fig:h2_bindcurve}}
	\end{figure}
	\begin{figure}
	\includegraphics[scale=0.525]{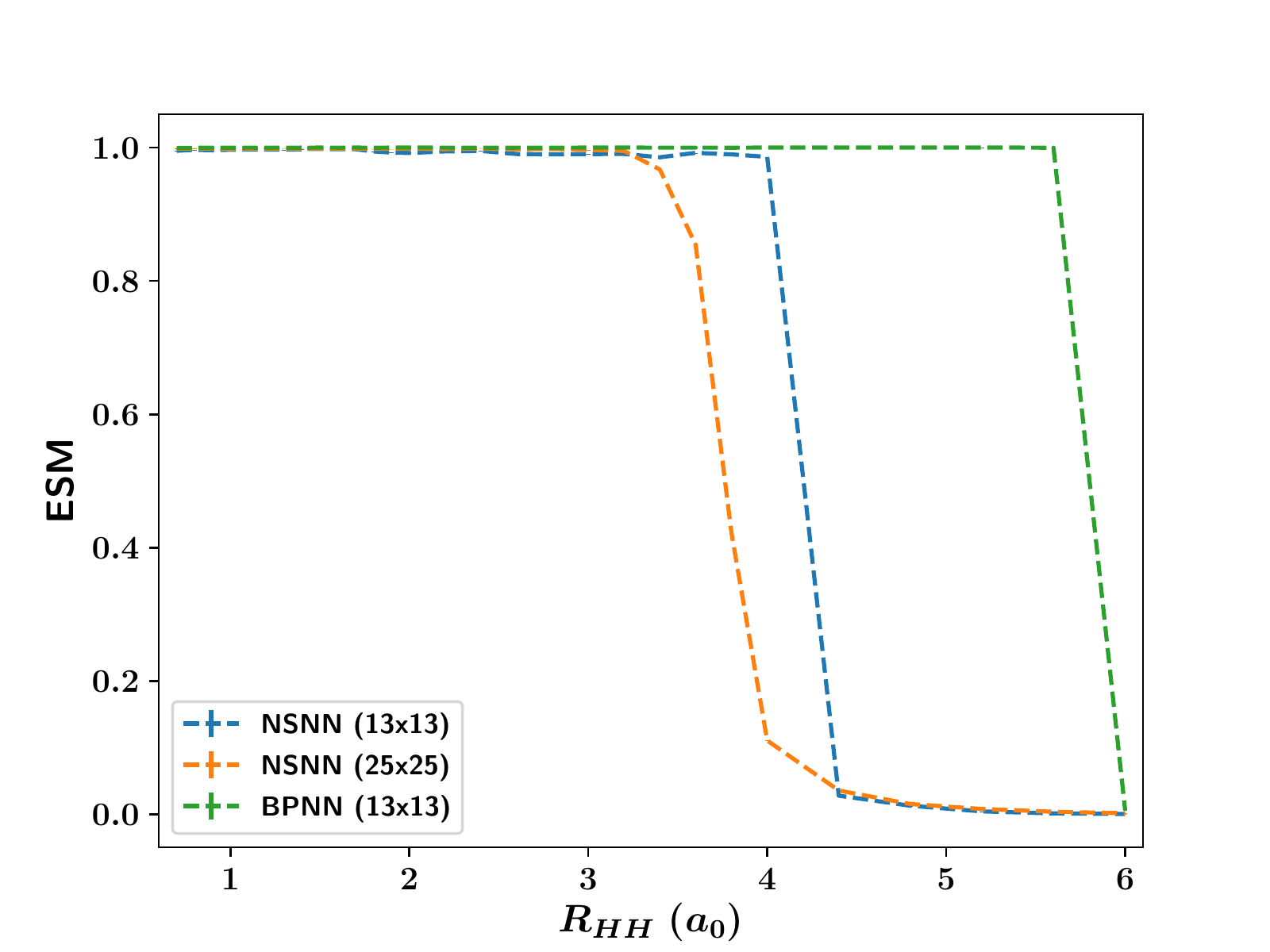}
	\caption{The ESM of our simple NNWF (NSNN) and simple bosonic product NNWF (BPNN) when employed to the H$_2$ molecule.
		\label{fig:h2_esm}}
	\end{figure}
	For the hydrogen dimer H$_2$, we considered both the simple NNWF (in two different sizes), as well as our simple bosonic product NNWF (as described in \ref{ssec:applications_h2}), to compute optimized energies for a range of protonic distances $R_{HH} = |\mathbf{R_1} - \mathbf{R_2}|$. The resulting energy curve $E_T(R_{HH})$ represents an approximation to the Born-Oppenheimer potential of the hydrogen dimer and is displayed in Fig.~\ref{subfig:h2_energies}, with the remaining differences to the exact potential energy curve $E_0(R_{HH})$ shown in Fig.~\ref{subfig:h2_residuals}. 
	The reference energy curve $E_0(R_{HH})$ is taken from Ref.~\onlinecite{pachucki_born-oppenheimer_2010} and extrapolated to the complete basis set limit, where for small distances a James-Coolidge basis is used, whereas Heitler-London functions with arbitrary polynomial in electron variables are employed for large distances. Whereas the simple NNWF underestimates the exact binding energy, this can be systematically improved by increase the size of the NNWF. The bosonic product NNWF, however, quantitatively reproduces the binding energy already for rather small sizes, which demonstrates its ability to accurately recover dynamic electron correlation effects. Nevertheless, despite the differences in accuracy, all of our NNWFs correctly predict the equlibrium distance to be at $R_{HH} = 1.4011\ a_0$. 
	
	However, the stretched H$_2$ molecule represents a rigorous challenge for even the most used accurate quantum-chemical approaches, such as two-particle reduced density matrix theory~\cite{PhysRevLett.93.213001,PhysRevLett.101.253002}, natural orbital functional theory~\cite{PhysRevLett.81.866,NOFT}, and density matrix renormalization group approaches~\cite{DMRG1,DMRG2}, to name just a few. This well-known problem is attributed to the multi-reference character of the stretched H$_2$ molecule, or static electron correlation that arises in situations with degeneracy or near-degeneracy, as in transition metal chemistry and strongly correlated systems in general~\cite{MoriCohen}. As a consequence, the stretched H$_2$ molecule and similar problems are typically dealt with using multi-determinant wave functions~\cite{MusialCC, MainzWIRES}. But, for larger systems with many degeneracies, the number of determinants quickly becomes unfeasible. In fact, due to the strong multi-reference character of the stretched H$_2$ molecule, the potential energy is substantially overestimated upon dissociation when using the simple NNWF. On the one hand, the error can be effectively suppressed in a systematic manner by increasing the size of the FFNN within the simple NNWF, as can be seen in Fig.~\ref{subfig:h2_residuals}. On the other hand, our bosonic product NNWF is able to reproduce the exact full configuration-interaction binding curve with great accuracy. This is to say that despite the strong multi-reference character of the strecthed H$_2$ molecule, the bosonic product NNWF is not only capable to recover the dynamic, but also the static correlation energy. 
	

	To complement the energy results, in Fig.~\ref{fig:h2_esm} the ESM diagnostics of Eq.~\ref{eq:esm} is shown for all employed NNWFs. Whereas all of our NNWFs successfully learned approximately exchange symmetric states for $R_{HH} < 3\ a_0$, they grasp asymmetric states starting from some distance $3\ a_0 < R_{HH} < 6\ a_0$, with the simple NNWFs becoming asymmetric earlier than the bosonic product NNWF. Just as for the soft-core harmonic trap system, the asymmetry stems from a localization of both particles on opposite sides of the system (here along the H-H axis), i.e. the NNWFs adopted a state of distinguishable particles.
	
	\section{Conclusions}
	
	To summarize, a novel VMC method using FFNN-based trial wave functions in continuous space is presented to approximate the ground states of model Hamiltonians to (in principle) arbitrary accuracy. We have formulated several versions of such FFNN-based trial wave functions, with different properties regarding their exchange symmetry and practicality. All NNWFs were tested on a correlated, but exactly solvable, 2-particle system, the soft-core harmonic trap. In all cases the exact ground state energies were predicted with good accuracy, in particular when considering the use of a relatively small FFNN. Nevertheless, we could show that including problem-specific knowledge within the NNWF construction offers potentially superior accuracy over a more simple, but general approach.
	This distinction in accuracy between the different NNWFs was more pronounced when for the $H_2$ molecule within the Born-Oppenheimer approximation.
	
	We observed that NNWFs without guaranteed exchange symmetry were prone to learn states of distinguishable particles whenever such a state was energetically close to the ground state of indistinguishable particles, though the practical consequences of this behavior remains indeterminate. Research on novel methods for providing exchange symmetric input to the FFNN, thereby preventing the asymmetry inherently, are currently underway and will be reported elsewhere. 
	
	In conclusion, we have demonstrated that our NNWFs represent a flexible trial wave function that can be straightforwardly applied to a variety of different Hamiltonians without necessarily requiring problem-specific adjustments. 
	We expect to achieve a significant speed-up of the optimization process by employing more suitable differentiation techniques to compute the necessary gradients, in particular by employing a reverse-mode differentiation scheme. However, our currently used forward-accumulation scheme will most likely remain an efficient choice for all second-order input derivatives, which are required to compute the exact kinetic energy of the NNWF. 
	
	Finally, it is worth noticing that the (anti-)symmetry requirement of the trial wave function can be considered as a serious limitation of our method, since the straightforward and general way of constructing (anti-)symmetric NNWFs via (anti-)symmetrization operators (see Eqs.~\ref{eq:psi_S} and \ref{eq:psi_AS}) is computationally not feasible for many-body applications. Fortunately, the fact that the lowest energy state is always bosonic makes it possible to simulate bosons by relying on the fact that the optimization will make the NNWF approximately symmetric. 
    When one is interested in the simulation of fermions, however, the problem becomes more serious. In fact, the approach we used in this work corresponds to a fixed-node approximation~\cite{PhysRevLett.45.566}. This means that the quality of our solution will depend on the nodal surface resulting from the choice of the functions that form the Slater determinant. 
    The possibility, to go beyond the fixed-node approximation and recover even more static electron correlation using a single Slater determinant only by means of the backflow transformation \cite{PhysRev.102.1189,PhysRevE.68.046707,ruggeri_nonlinear_2018, PhysRevLett.122.226401}, shadow wave function \cite{PhysRevLett.60.1970, PhysRevB.38.4516, PhysRevE.90.053304, calcavecchia_epl, ZFNA}, or resonating valence bond approach \cite{RVB, O3RVB, TurboRVB}, is to be underlined and will be presented elsewhere.
	
	\acknowledgements
	The generous allocation of computing time on the FPGA-based supercomputer ``Noctua'' by the Paderborn Center for Parallel Computing (PC$^2$) is kindly acknowledged. This project has received funding from the European Research Council (ERC) under the European Union's Horizon 2020 research and innovation programme (grant agreement No 716142).

\section*{Data Availability Statement}
The data that support the findings of this study are available from the corresponding author upon reasonable request.

	\appendix
	
	\section{Appendix}
	\subsection{Activation Functions}
	To investigate the sensitivity of the chosen activation function on the eventually achieved accuracy, besides the $a_T$ and $a_G$ hidden layer activation functions mentioned in section \ref{ssec:ffnn}, we also explored the identity function for some of the hidden units, i.e.
	\begin{equation}
		a_I(p) = p.
		\label{eq:id_actf}
	\end{equation}
	Furthermore, in addition to an exponential activation function for the output unit, we also tried the commonly employed logistic activation function
	\begin{equation}
    	a_L(p) = \frac{1}{1+\exp(-p)},
    	\label{eq:log_actf}
	\end{equation}
	which is bounded between 0 and 1. Note that this range restriction is not a problem in our VMC method, as the scale or normalization of the wave function is not relevant.
	
	We dismissed, however, the common choice of linear output activation functions, since in our testing we observed unfavorably larger oscillations around zero for input values, where the wave function has a small amplitude. Moreover, in the bosonic case, we want to avoid a wave function with variable sign anyway.
	
\begin{figure}
		\includegraphics[scale=0.5]{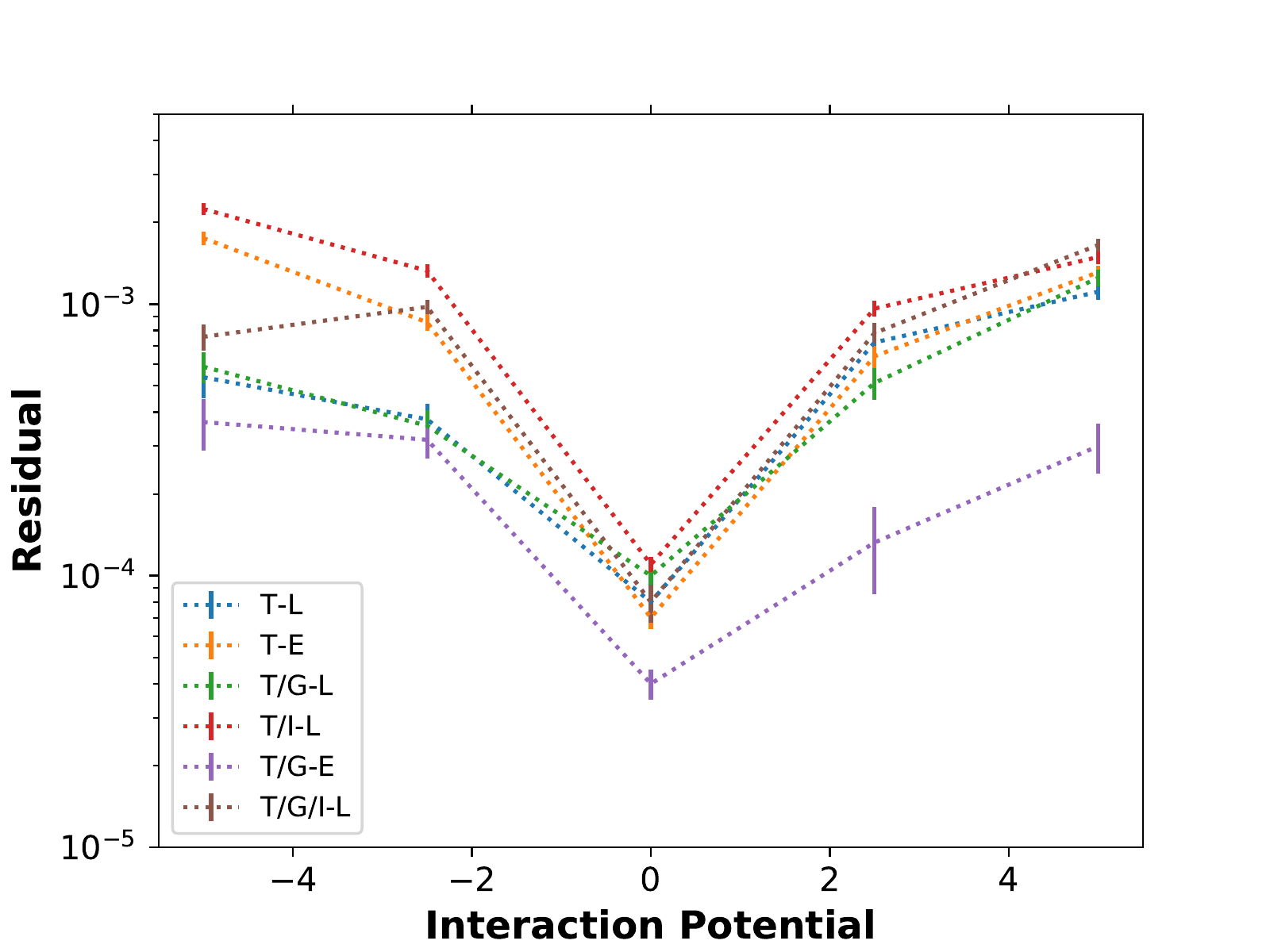}
		\caption{The energy residuals with respect to the exact ground state energies for the simple NNWF applied to the bosonic harmonic soft-core trap system. Results for different combinations of hidden layer and output activation functions are displayed, where $T$ refers to hyperbolic tangent sigmoid, $G$ to Gaussian, $I$ to the identity, $L$ to logistic and $E$ to exponential activation functions, respectively. \label{fig:comp_actfs}}
	\end{figure}

\begin{figure}
		\includegraphics[scale=0.5]{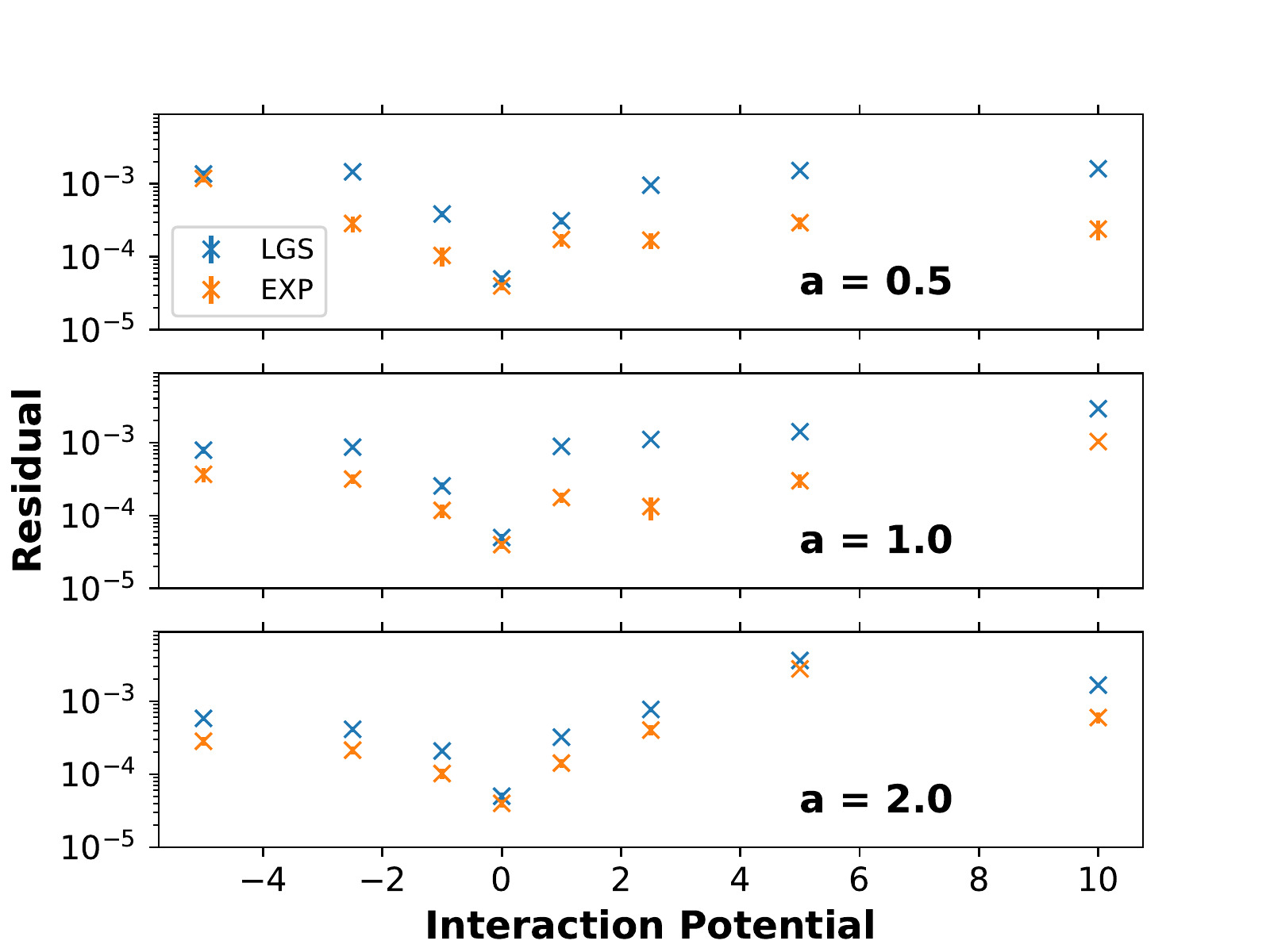}
		\caption{The energy residuals with respect to the exact ground state energies for the simple NNWF applied to the bosonic harmonic soft-core trap system. Results for T/G-L (LGS) and T/G-E (EXP) as hidden layer and activation functions are shown for different Hamiltonian parameters, where $T$ refers to hyperbolic tangent sigmoid, $G$ to Gaussian, $L$ to logistic and $E$ to exponential activation functions, respectively. 
		\label{fig:comp_output_actfs}}
	\end{figure}

	The energy residuals results for the harmonic soft-core trap togther with selected combinations of hidden layer and output activation functions are dis`played in Fig.~\ref{fig:comp_actfs}. The corresponding results for the combination of $a_T$ and $a_G$ hidden layer activation function with either $a_L$ or $a_E$ on the output, respectively, are shown in Fig.~\ref{fig:comp_output_actfs}.
	
	\subsection{Hidden Layer Structure}
\begin{figure}
		\includegraphics[scale=0.5]{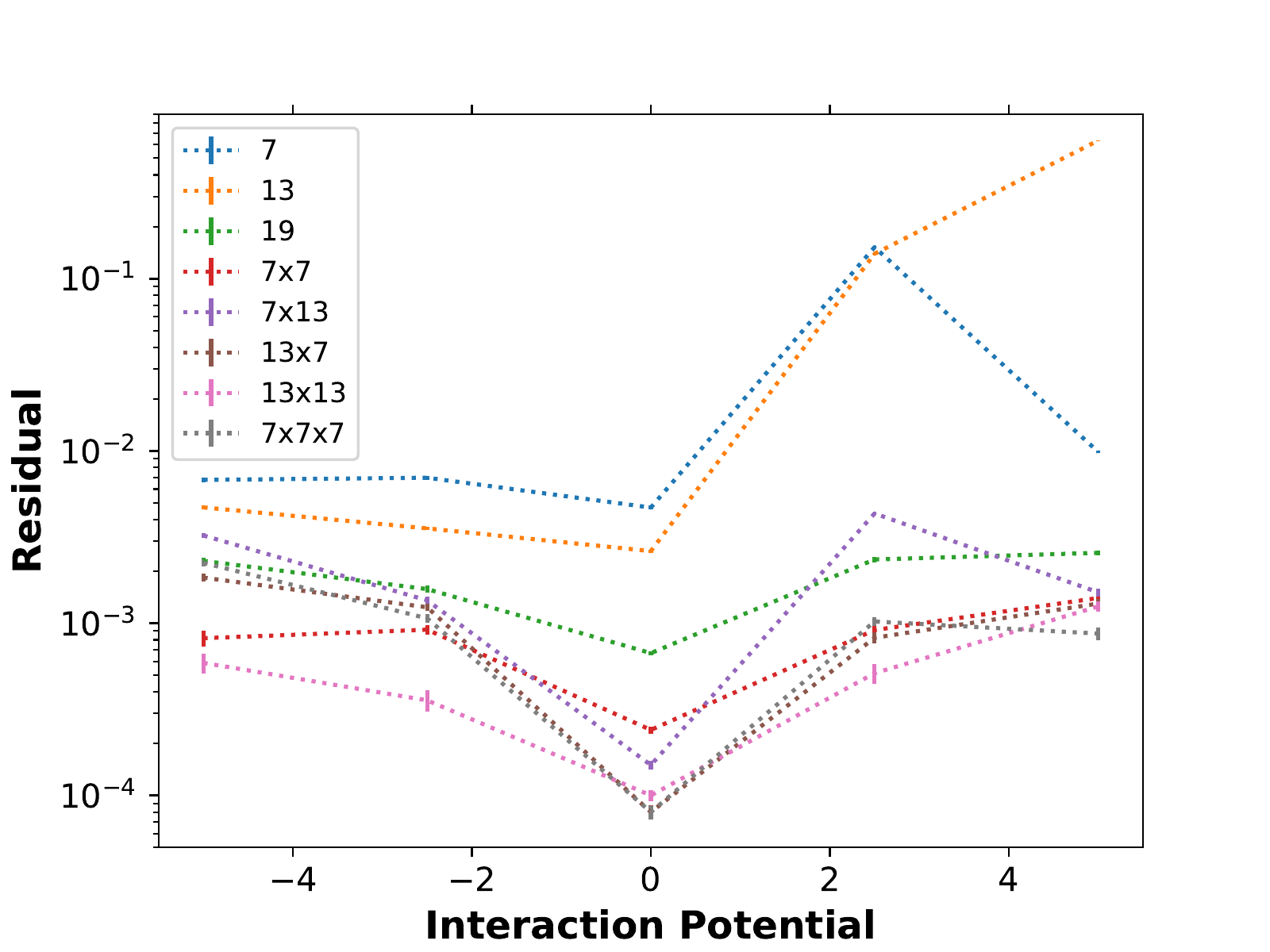}
		\caption{The energy residuals with respect to the exact ground state energies for the simple NNWF applied to the bosonic harmonic soft-core trap system. Results for the T/G-E activation function configuration is shown as a function of different number of hidden layers and hidden units, where $T$ refers to hyperbolic tangent sigmoid, $G$ to Gaussian and $E$ to exponential activation functions, respectively. 
		\label{fig:comp_units}}
	\end{figure}

	Similarly, we also investigated how the accuracy changes when changing the size and number of hidden layers. An overview of our results is shown in Fig.~\ref{fig:comp_units}.

    \bibliography{references}

\end{document}